\documentclass{aa}  

\usepackage{graphicx}
\usepackage{txfonts}
\usepackage{xcolor}
\usepackage{amsmath}
\usepackage{amssymb}
\usepackage{graphicx}
\usepackage{dblfloatfix} 

\usepackage{hyperref}

\begin{document} 

   \title{Stellar rotation in the intermediate-age massive cluster NGC~1783: clues on the nature of UV-dim stars\thanks{Based on observations collected at the European Southern Observatory under ESO programme 111.24LZ.001 (PI: Dalessandro).}}

   \author{S. Leanza \inst{1,2} \and
          E. Dalessandro \inst{2} \and
          M. Cadelano \inst{1,2} \and
          C. Fanelli \inst{2} \and
          G. Ettorre \inst{1,2} \and
          S. Kamann \inst{3} \and
          N. Bastian \inst{4,5} \and
          S. Martocchia \inst{6,7}\and
          M. Salaris \inst{3} \and
          C. Lardo  \inst{1,2} \and 
          A. Mucciarelli \inst{1,2} \and
          S. Saracino \inst{3,8}} 

   \institute{Department of Physics and Astronomy ‘Augusto Righi’, University of Bologna, via Gobetti 93/2, I-40129 Bologna, Italy
    \and
    INAF – Astrophysics and Space Science Observatory of Bologna, via Gobetti 93/3, I-40129 Bologna, Italy
     \and
     Astrophysics Research Institute, Liverpool John Moores University, IC2 Liverpool Science Park, 146 Brownlow Hill, Liverpool L3 5RF, UK
     \and
     Donostia International Physics Center (DIPC), Paseo Manuel de Lardizabal, 4, 20018, Donostia-San Sebasti\'an, Guipuzkoa, Spain
     \and
     IKERBASQUE, Basque Foundation for Science, 48013, Bilbao, Spain 
     \and
     Astronomisches Rechen-Institut, Zentrum für Astronomie der Universität Heidelberg, Mönchhofstraße 12-14, D-69120 Heidelberg, Germany
     \and 
     Aix Marseille Université, CNRS, CNES, LAM, Marseille, France
     \and 
     INAF – Osservatorio Astrofisico di Arcetri, Largo E. Fermi 5, 50125 Firenze, Italy }

   \date{}

 
\abstract{
Over the past decade, stellar rotation has 
emerged as a key factor in shaping the morphology of color-magnitude diagrams of young and intermediate-age star clusters.
In this study, we use MUSE integral-field spectroscopy to investigate the stellar rotation of $\sim2300$ stars in the 1.5 Gyr old cluster NGC 1783 in the Large Magellanic Cloud. 
The effective temperature, surface gravity, radial velocity, and projected rotational velocity ($v\mathrm{sin}i$) of the entire sample
were obtained within a Bayesian framework to derive robust estimates of these parameters along with their associated errors.
The analysis shows that stars along the extended main sequence turn-off (eMSTO) cover a wide range of rotational velocities, from values consistent with no/slow rotation up to $v\mathrm{sin}i\sim250$ km s$^{-1}$. 
The distribution of stellar rotation velocities appears to play a crucial role in explaining the broadening of the eMSTO in this cluster, 
and a correlation is observed between
$v\mathrm{sin}i$ and the color of the eMSTO stars, with $v\mathrm{sin}i$ increasing as the color becomes redder. 
Among the eMSTO stars, we investigate the peculiar population of stars strongly dimmed in the UV (so-called UV-dim stars), recently discovered in NGC 1783.
UV-dim stars show clear photometric evidence of self-extinction and mild spectroscopic signatures typically observed in shell stars, thus suggesting that they have likely a decretion disc observed nearly equator-on.
Interestingly, the study
also shows that a significant fraction of UV-dim stars are slow rotators. We discuss potential implications these results may have on our understanding of the formation and evolution of UV-dim stars and we propose
that the rotational properties of the UV-dim stars should vary with cluster age.  
}


   \maketitle
%
\section{Introduction}
The Magellanic Clouds host a sizable population of young and intermediate-age (less than a few Gyrs) massive ($\sim10^5 M_{\odot}$)
stellar clusters, which are instead largely absent in the Milky Way. These clusters are the only recently formed massive stellar systems close enough to be resolved in individual stars. Therefore, they represent ideal laboratories for constraining the physical mechanisms at the basis of cluster formation, studying their early evolution with a level of detail that cannot be achieved for distant systems, and understanding the complex stellar evolutionary properties of relatively young stellar populations. 

It is now well established that the color-magnitude-diagrams (CMDs) of young- and intermediate-age stellar clusters display features that are not well reproduced through simple single stellar models.
One of the most striking aspects that characterizes virtually all clusters younger than $\sim 2$ Gyr is that they show extended main sequence turn-off regions (hereafter eMSTOs; \citealt{Mackey&Broby+07,Girardi+09,Goudfrooij+11,Goudfrooij+14,correnti+14,milone+15,milone+23a,Kamann+20,kamann+23})
and clusters younger than $\sim 500$ Myr also exhibit
split/dual main sequences \citep[see e.g.,][]{milone+16, milone+17, milone+23a}. Interestingly, these phenomena are observed also in lower-mass Galactic open clusters
\citep[e.g.,][]{Marino+18, Ovelar+20, Cordoni+24}, suggesting that the underlying mechanisms work irrespective of cluster masses, metallicity, and birth environments.

The presence of eMSTOs and split MSs has traditionally been interpreted as due to stars that formed at different epochs within the parent cluster, with age spreads of 150-500 Myr
\citep[e.g.,][]{Girardi+09,Goudfrooij+11,Goudfrooij+14,correnti+14}.
However, recent photometric and extended spectroscopic studies of stars along the eMSTOs of Magellanic Clouds massive clusters and open clusters in the Milky Way \citep[e.g.,][]{Bastian+18,marino+18b,Kamann+20,kamann+23,Bodensteiner+23,Dresbach+23} 
have demonstrated a direct correlation between the color of stars in the MSTO region and their projected equatorial velocity ($v\mathrm{sin}i$). These studies reveal that rotational velocity increases from blue to red, reaching values of up to approximately $250$ km s$^{-1}$. 
Indeed, rotation modifies the internal structure of stars and influences their evolution in several ways \citep[see, e.g.,][]{Meynet+20}.
The centrifugal support and rotationally induced mixing in the stellar interior may alter the hydrostatic equilibrium and the hydrogen content in the core of the star compared to non-rotating stars of the same mass and composition.
This results in a different evolutionary path of the rotating stars, leading to changes in their MS lifetimes and, consequently, their positions in the CMD with respect to the non-rotating stars \citep[see e.g.,][]{Wang+23}.
Furthermore, (fast) rotation can make stars oblate, and the resulting loss of spherical symmetry leads to effective temperature variations across their surface. In particular, rotation causes the so-called gravity darkening \citep{vonZeipel_1924}, as the centrifugal force reduces the effective gravity at the equator, the equatorial regions became dimmer and cooler than the polar regions. As a result, the observed magnitude and color of a rotating star become a function of its angular velocity and of 
the inclination of its rotation axis towards the line of sight, with a star seen pole-on appearing bluer and brighter than the same star observed equator-on at a fixed spin velocity.

The physical mechanisms responsible for the distribution of rotation rates ($0-250$ km s$^{-1}$; e.g., \citealt{kamann+23}) needed to reproduce the observed morphology of the eMSTOs is still a matter of debate. For example, \citet[][see also \citealt{Abt+04}]{d'antona+15} 
suggested that binary stars can play a role. In this framework the idea is that
all stars were born with relatively high rotational velocities, 
with a subset subsequently slowing down due to the tidal torques in binary systems. 
However, based on large and multi-epoch spectroscopic observations it has been recently shown (e.g., \citealt{Kamann+20,kamann+21}) that there is no significant difference 
in the binary fraction of 
the red and blue main sequence stars,
thus possibly suggesting that binarity is not a dominant mechanism in the formation of the observed bimodal rotational distributions.
On the other hand, the presence of unresolved binaries may shift the positions of stars along the main sequence, potentially blurring the expected correlation between the color of the stars along the eMSTO and their rotational velocities.
Alternatively, the observed rotation distributions have been suggested to be the result of the difference in the lifetimes of stars's pre-MS disks \citep{bastian+20} or of stellar mergers \citep[e.g.,][]{wang+22}.

\citet{milone+22} and \citet{milone+23b} have recently unveiled the existence of an additional peculiarity on the eMSTOs of intermediate and young clusters in the Magellanic Clouds. In \citet{milone+23b}, in a selection of clusters younger than $\sim200$ Myr, they discovered a group of bright MS stars that are distributed at significantly redder colors than the MSTO mean locus in CMDs involving the F225W and F275W UV bands. \citet{milone+23b} observed that these stars overlap with the other eMSTO stars in optical color combinations and therefore they argued that UV-dim stars are on average slow-rotators.
However, arguments based on the distribution and fraction of UV-dim stars in the CMDs of massive clusters with different ages \citep[$<2$ Gyr;][]{martocchia+23} have been used to suggest that, contrary to previous assumptions, these stars are more likely to be fast rotators.
Interestingly, these results have been directly confirmed by the direct rotational velocity measurements of eMSTO stars in the very young massive cluster NGC~1850 by \citet{kamann+23}.
In particular, the spectroscopic analysis of MUSE data by \citet{kamann+23} suggests that the UV-dim stars in NGC~1850 are Be stars (i.e. fast rotating B stars with decretion disks) observed almost equator-on, such that they are seen through their excretion disks \citep[e.g.,][]{Rivinius+06} and they are thus self-extincted. This is also supported by the shell-like features observed in these stars \citep{kamann+23}.
However, it is still unclear whether the same mechanism can be applied to lower-mass stars in intermediate-age clusters, in part because the disk properties of A-type stars are more difficult to study because these stars do not ionize their disks.
Along similar lines, based on a large suite of simulations, \citet{d'antona+23} have recently pointed out that self-extinction due to the presence of a disk can in principle reproduce the observed width of eMSTOs in young- intermediate-age clusters.  

In this paper we study the relevant aspects related to the impact of stellar rotation on the distribution of stars in the CMDs of young and intermediate-age clusters, by targeting the $~1.5$ Gyr old 
\citep{mucciarelli+07,Zhang_2018} and massive 
\citep[$2\times10^5 M_{\odot}$;][]{song+21} stellar cluster NGC~1783 in the Large Magellanic Cloud (LMC). This system represents an ideal target in this context as it is located in a region of the LMC characterized by low extinction ($A_V<0.1$ mag) and by a negligible field contamination. 
The paper is structured as follows. In Section~\ref{sec:obs} the adopted data set is described while in Section~\ref{sec:pampelmuse} we present the data-reduction procedure. The analysis methodology is presented in Section~\ref{sec:spectral_fit}. In Section~\ref{sec:membership} we introduce the final data sample. 
Results are reported in Section~\ref{sec:results} and Section~\ref{sec:uvdim} focuses on UV-dim stars. In Section~\ref{sec:discussion} we summarize the main findings and discuss their implications.

\begin{figure}[b]
\centering
\includegraphics[width=0.52\textwidth]{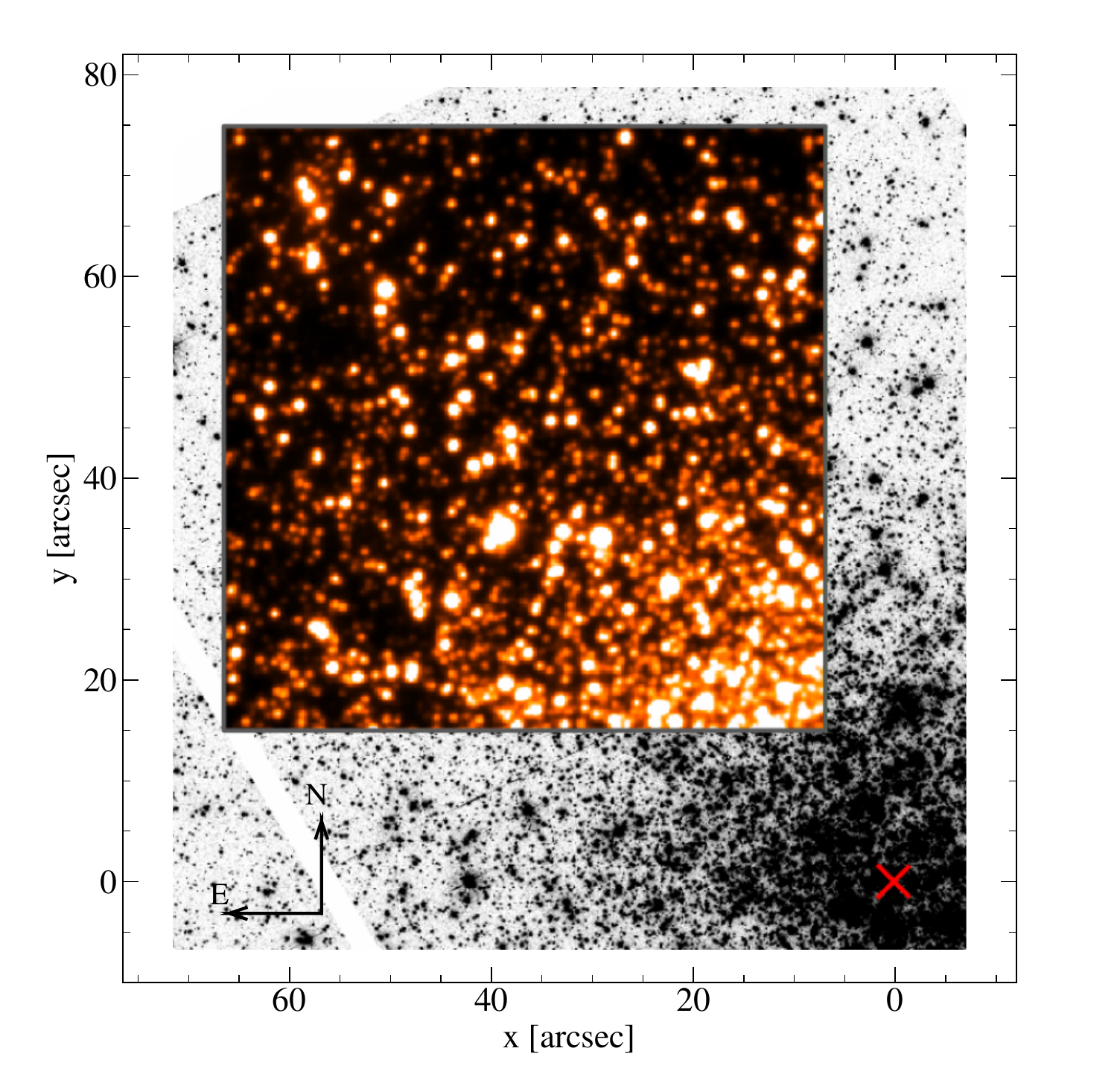}
\centering
\caption{Stacked image created from the combined MUSE/WFM datacube
obtained for NGC 1783 in this work (see Section \ref{sec:obs}). For reference, we show the HST/ACS image in the F555W filter (colored in gray-scale) in the background, while the cluster center is marked by the red cross.}
\label{fig:fov}
\end{figure}

\section{Observations}
\label{sec:obs}
The observational data-set used in this work was obtained with the integral-field spectrograph MUSE@ESO-VLT \citep{bacon+10} in the wide-field mode (WFM) configuration. 
MUSE/WFM covers a field of view of $1\arcmin\times 1\arcmin$ with a spatial sampling of $0.2\arcsec$/pixel. The data were collected in the nominal mode, which provides a wavelength coverage from $4750$ \AA\ to $9350$ \AA\ and a
resolving power ranging between $1800$ in the blue part and $3500$
in the red part of the spectral range.
The data set consists of 10 MUSE acquisitions of NGC 1783 obtained between September and November 2023 (program: 111.24LZ.001, PI: Dalessandro).
All pointings target a region of the cluster that is slightly offset from the center by about $10\arcsec$ towards the northwest direction (see Fig.~\ref{fig:fov}), to avoid contamination effects due to the overcrowding of the central regions.
Each acquisition consists of four exposures of $\sim$660 s integration time each, executed applying a derotator offset of $90\degr$ and a
small dithering pattern between consecutive exposures to remove possible resolution differences between the individual spectrographs, for a total of 40 independent exposures. 

The data reduction was performed using the standard  MUSE pipeline \citep{Weilbacher+20} through the EsoReflex environment \citep{Freudling+13}.
The pipeline first performs the basic reduction (bias subtraction, ﬂat-ﬁelding, and wavelength calibration) for each individual integral-field unit (IFU). Then the sky subtraction is applied, all the preprocessed data are converted into physical quantities by
performing the ﬂux and astrometric calibrations, and the heliocentric 
velocity correction is applied.
Afterwards, the data from all 24 IFUs are combined in a single datacube for each exposure.

To maximize the signal-to-noise ratio (S/N) of all sources in the observed field of view, we combined all the available observations into a single datacube, by taking into account dithering offsets and rotations among different exposures, for a total integration time of $\sim$8 hours.
As a last step, we removed residual sky contamination from the MUSE datacube using the dedicated workflow included in ESO reflex, \texttt{use$\_$zap.xml}, which is based on the Zurich Atmosphere Purge code \citep[ZAP,][]{soto+16}.
In Fig. \ref{fig:fov} we show the stacked image of the MUSE datacube created from the combination of all the available observations.

In the MUSE datacube, we observe a significant diffuse emission from a close star forming region. 
This emission is evident in several spectral lines, including $\lambda4861$ (H$\beta$), $\lambda5007$ (O~III), $\lambda6563$ (H$\alpha$), $\lambda6584 \AA$ (N~II), and $\lambda6716$ and $\lambda6731$ (S~II).
Unfortunately, these emission lines are also visible in the MUSE spectra of individual stars, particularly affecting the spectra of fainter stars. This is especially true for the H$\alpha$ line, making it unusable in our analysis.

\section{Extraction of the spectra}
\label{sec:pampelmuse}
Stellar spectra were extracted from the combined MUSE datacube by using \texttt{PampelMuse} \citep{kamann+13}. This is a software optimized for the deblending 
of resolved sources in integral-field spectroscopic data. 
The code uses an astrometric reference catalog to determine the position of each star in the MUSE datacube and to fit the sources from the reference catalog to the data by using a wavelength-dependent MUSE point spread function (PSF) model.
Based on this information, \texttt{PampelMuse} extracts the deblended spectra of all the resolved stars in the MUSE field of view.
In our analysis, we used as astrometric reference catalog the \textit{Hubble Space Telescope} (HST) multi-band photometric catalog presented in \citet{Cadelano+22}.

\begin{figure}[ht!]
\centering
\includegraphics[width=0.42\textwidth]{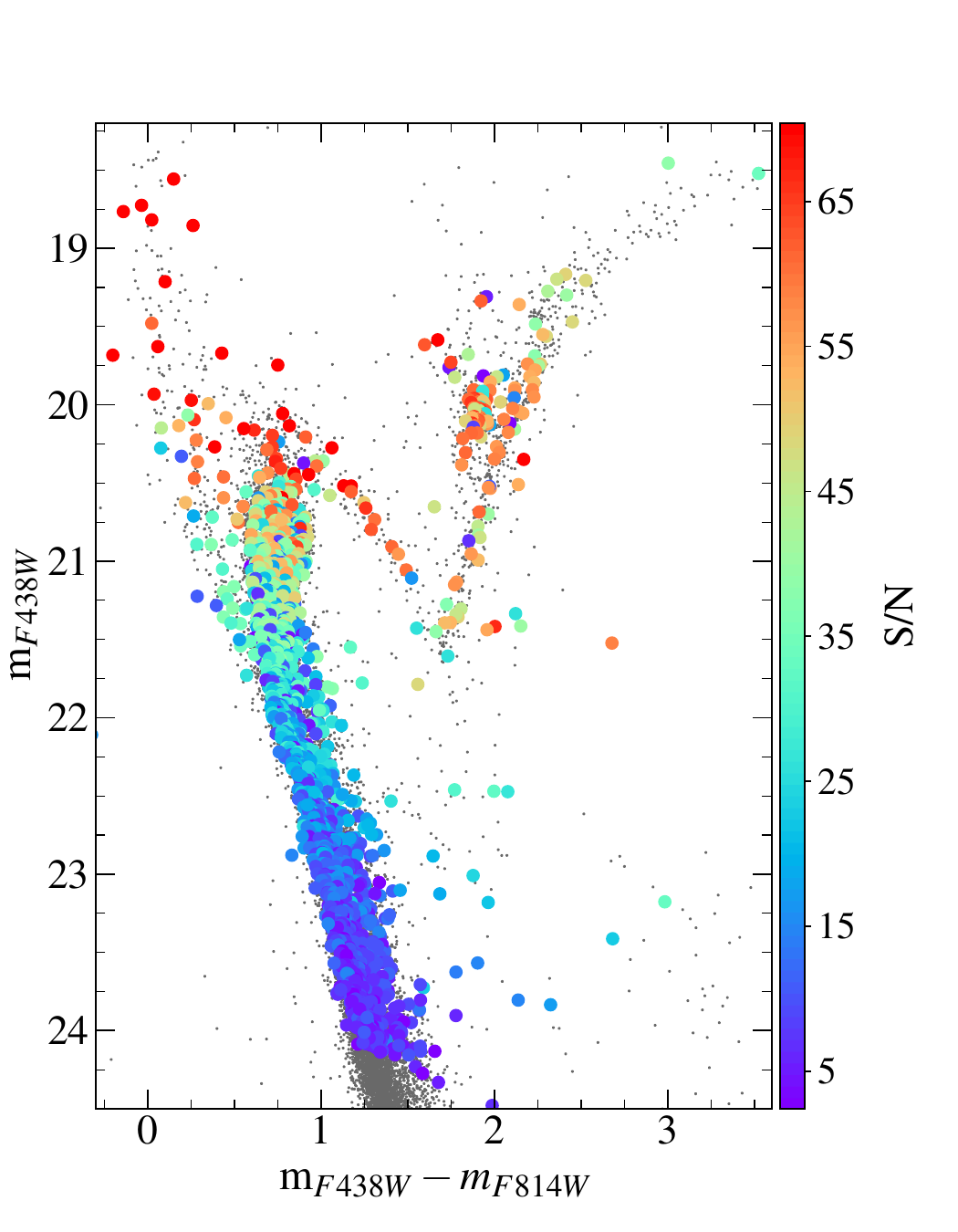}
\centering
\caption{CMD of NGC 1783 produced from the HST photometric
  catalog described in \citet[][gray dots]{Cadelano+22}. 
  The MUSE targets for which useful spectra were extracted from the combined datacube are shown by the large circles color-coded according to the spectral S/N.}
\label{fig:cmd_snr}
\end{figure}

The extraction procedure yielded a total of 2293 spectra of individual stars, 
for which the deblending algorithm was able to extract the contribution of a single source.
Fig. \ref{fig:cmd_snr} shows the distribution of the obtained sample of stars in the (m$_{\rm F438W}$, m$_{\rm F438W}-$m$_{\rm F814W}$) CMD of NGC 1783 color-coded according to the S/N of the spectra. 
The extracted spectra sample a wide magnitude range from 
$m_{F438W}$ $\sim$ 18 down to a magnitude threshold 
of $m_{F438W}$ $\sim$ 24, about 3 magnitudes below the MSTO. 
Also, we note that the distribution of the extracted spectra guarantees an excellent sampling of the eMSTO
of NGC 1783 with $\sim 600$ extracted stars in the magnitude range 
$20.5 \leq m_{F438W} \leq 22.0$,  and
with S/N typically larger than $\sim30$.

\section{Spectral fitting}
\label{sec:spectral_fit}
To analyze the MUSE spectra we developed an algorithm based on a Bayesian approach to determine the best-fit stellar parameters of individual stars.
This method allows the simultaneous estimate of the effective temperature ($T_{\mathrm{eff}}$),
surface gravity ($\mathrm{log} \ g$), radial velocity (RV), and 
projected rotational velocity ($v\mathrm{sin}i$) of the stars by means of a direct comparison between the observed spectra and
a library of synthetic spectra exploring reasonable ranges of values for each parameter. 
We generated a library of synthetic spectra through the code SYNTHE
(\citealt{sbordone+04} and \citealt{kurucz+05}) using the ATLAS9 atmosphere model
\citep{kurucz+05}.
The grid of synthetic spectra was computed assuming values of $T_{\mathrm{eff}}$ 
ranging from 3500 K to 8800 K, in steps of 50 K,  
and of $\mathrm{log} \ g$ ranging from 0.4 to 5.0, in steps of 0.1. 
This fine and uniform sampling of the atmospheric parameters ensures 
a comprehensive coverage of all the evolutionary stages sampled by the observed targets, 
thus avoiding the introduction of bias in the subsequent fitting procedure.
Synthetic spectra were obtained by adopting a solar [$\alpha$/Fe] abundance ratio and the cluster metallicity [Fe/H] $=-0.35$ dex  \citep{mucciarelli+08}. Finally, the spectra were computed in the wavelength range covered by MUSE ($4900-9300$ \AA ),
with similar spectral resolution (R$=$3000) 
by convolving with a Gaussian profile,
and the same spectral sampling (0.125 nm$/$pixel).
\\
To compare each observed spectrum with the synthetic spectra,
we used a Markov Chain Monte Carlo (MCMC) sampling technique. This approach provides also robust uncertainty estimates for all output parameters.
The fitting algorithm can be schematically described as follows.
\begin{enumerate}

\item As a first step, the observed spectra were normalized to the continuum. Then for each spectrum the S/N was estimated as the ratio between the average of the counts and its standard deviation in the 8000 $-$ 9000 \AA \ wavelength range. Finally, the spectral region to use for the following fit was selected.
In particular, instead of using the full MUSE spectrum, we applied the fitting procedure to the wavelength range 7750 $-$ 8900 \AA. This range was chosen as it contains the most prominent spectral features at the MUSE resolution, namely the Ca II triplet lines and the hydrogen Paschen lines,
and to avoid regions strongly affected by residuals of telluric absorption bands, which can potentially lead to mismatches between the observed and synthetic spectra.
We note that the selected spectral region is particularly suitable for our aims because the Ca II triplet and Paschen lines are sensitive to variations in $T_{\mathrm{eff}}$ (especially the Paschen lines) and $\mathrm{log} \ g$. These lines are also ideal for measuring RVs and rotational velocities from line broadening along the line of sight, since they are well-defined even at lower S/N. 
We have verified that the procedure gives consistent results across different stellar parameters (e.g., $T_{\mathrm{eff}}$, $\mathrm{log} \ g$ and rotational velocity) when using the full spectrum masking the strong telluric absorption band regions.
However, since the use of the full spectrum is more computationally and time expensive and our data set includes a large number of targets, the results presented in the following are obtained by only using the selected portion of the spectrum.


\item In the second step uniform priors are defined for $T_{\mathrm{eff}}$, $\mathrm{log} \ g$, RV, and $v\mathrm{sin}i$ across a wide range of values: $T_{\mathrm{eff}}$ from 3500 to 8800 K, $\mathrm{log} \ g$ from 0.4 to 5.0, RV from 70 to 470 km s$^{-1}$, and $v\mathrm{sin}i$ from 0 to 300 km s$^{-1}$, ensuring a thorough exploration of the parameter space.
During each MCMC iteration, the parameters are sampled from these prior distributions, starting with random values as first guesses, 
and the synthetic spectrum with the nearest $T_{\mathrm{eff}}$ and $\mathrm{log} \ g$ is selected from the grid.
The rotational broadening, corresponding to the sampled value of $v\mathrm{sin}i$,  is applied to the selected synthetic spectrum using the \texttt{PyAstronomy.pyasl.rotBroad}\footnote{\url{https://github.com/sczesla/PyAstronomy}} \citep{pyalsRot} Python function. 
The synthetic spectrum is then shifted in wavelength based on the Doppler effect according to the sampled value of RV.
Finally, the similarity between the resulting synthetic and the observed spectrum is quantified by computing the following logarithmic likelihood function within the spectral range selected in the step 1:

\begin{equation}
  \mathrm{ln}\mathcal{L}=-0.5\sum_{i}^{}\Big[\frac{(f_{obs,i}-f_{syn,i})^2}{\sigma_i^2}\Big]
\end{equation}

where $f_{obs,i}$ and $f_{syn,i}$ are the fluxes of the $i^{th}$ spectral pixel in the selected 
wavelength range  (7750 $-$ 8900 \AA) of the observed and synthetic spectra, respectively. While $\sigma_i$ is the uncertainty of the observed flux at the $i^{th}$ spectral bin, computed as the quadratic sum of the inverse of the S/N and the variance of the flux of the corresponding spectral pixel\footnote{The variance values are provided by the MUSE data reduction pipeline.}.

The algorithm searches for the best-fit synthetic model -- i.e. the combination of parameters that maximizes the previous likelihood function -- by sampling the parameter space ($T_{\mathrm{eff}}$, $\mathrm{log} \ g$, RV, and $v\mathrm{sin}i$) using the \texttt{emcee} code \citep{Foreman+13}.
As a result of the procedure, the best-fit values for $T_{\mathrm{eff}}$, $\mathrm{log} \ g$, RV, and $v\mathrm{sin}i$ are obtained, together with the associated uncertainties corresponding to the 50th, 16th and 84th percentiles of the posterior distributions, respectively.
\end{enumerate}

\subsection{Validation of the method}
\label{sec:simu}
\begin{figure}[h]
\centering
\includegraphics[width=0.45\textwidth]{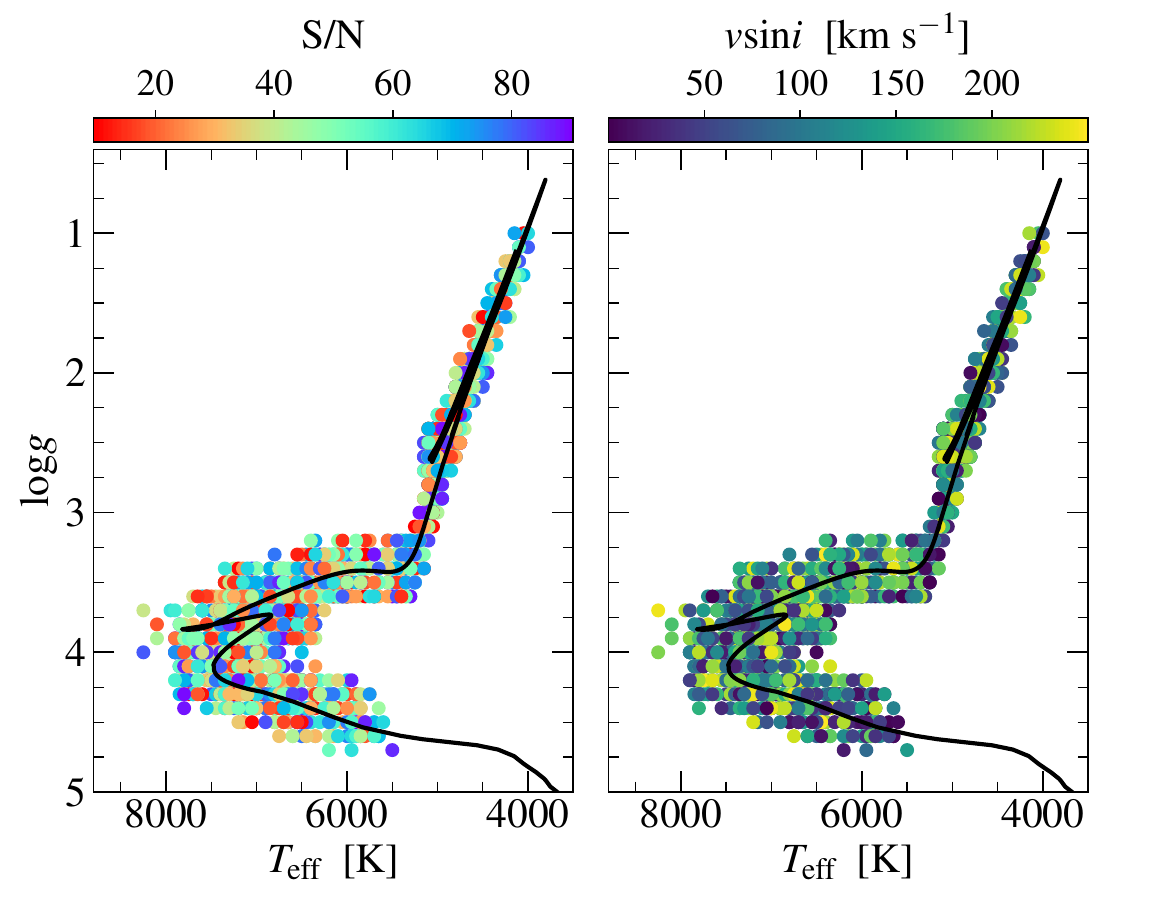}
\centering
\caption{The panels show the position of the simulated targets (large circles) in the $\mathrm{log} \ g$-$T_{\mathrm{eff}}$ plane color-coded according to the S/N and
rotational velocity ($v\mathrm{sin}i$) of the spectra in the left and right panels, respectively. The black curve is the isochrone 
computed for a stellar population of 1.5 Gyr and [Fe/H] $=-0.35$.}
\label{fig:data_simu}
\end{figure}

To validate the accuracy and reliability of the adopted method, we tested the procedure on a set of simulated spectra.
To construct the simulated data set, we generated 2000 pairs of $T_{\mathrm{eff}}$ and $\mathrm{log} \ g$ values using a BaSTI-IAC isochrone \citep{Hidalgo+18} of appropriate age (1.5 Gyr) and metallicity ([Fe/H] $=-0.35$) for NGC 1783, as a reference to ensure a close match to the observational data set.
Then, for each  $T_{\mathrm{eff}}$-$\mathrm{log} \ g$ pair, we extracted the corresponding spectrum from the library of synthetic models described in Section \ref{sec:spectral_fit}. 
The RV was set to zero for all simulated targets, while each spectrum was broadened to a certain rotational velocity with random values ranging from 0 to 250 km~s$^{-1}$. 
We also added random amounts of Poisson noise to the selected synthetic models to simulate spectra with S/N ratios ranging from 10 to 90 to match the observations.
The position in the $\mathrm{log} \ g$-$T_{\mathrm{eff}}$ diagram of the resulting simulated data set is shown in Fig. \ref{fig:data_simu}, 
where the color scale represents the S/N and $v\mathrm{sin}i$ values in the left and right panels, respectively.
\begin{figure}[h!]
\centering
\includegraphics[width=0.38\textwidth]{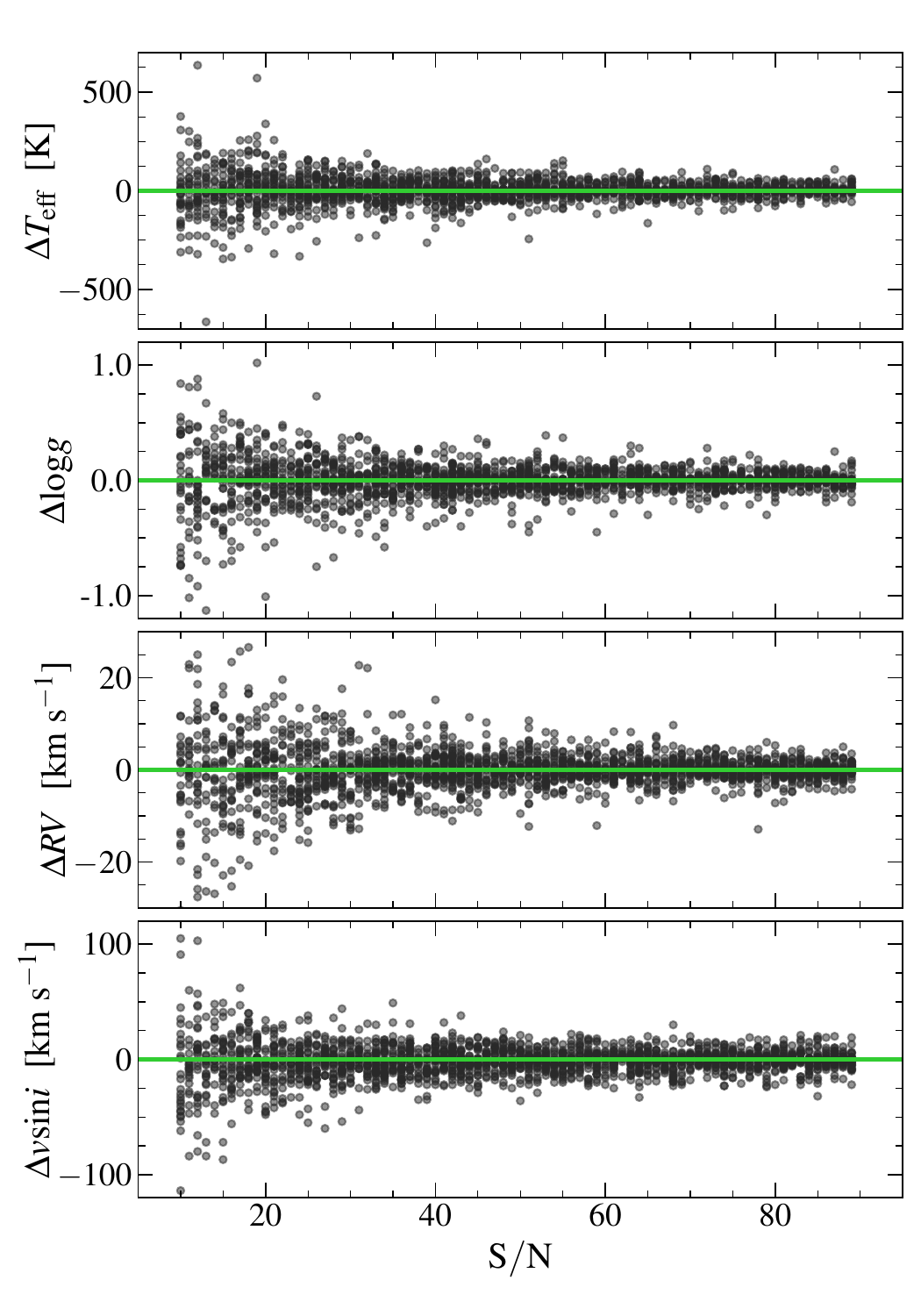}
\centering
\caption{Difference between the parameter values recovered by the fitting algorithm and the known input values as a function of the S/N of the simulated spectra, for $T_{\mathrm{eff}}$, $\mathrm{log} \ g$, RV, and $v\mathrm{sin}i$ from top to bottom panels.}
\label{fig:simu_diff}
\end{figure}

We applied the spectral fitting method described in the previous section to the simulated data to verify that the procedure is able to recover the input values of the parameters 
without introducing bias,
and to test the effectiveness and the robustness of the adopted MCMC algorithm.
The results obtained are shown in Fig. \ref{fig:simu_diff} as the distribution of the difference between the best-fit parameters resulting from the fitting procedure and the input values, as a function of the S/N of the simulated spectra.
Fig. \ref{fig:simu_diff} shows that all distributions are centered on zero at all S/N values, meaning that the values of $T_{\mathrm{eff}}$, $\mathrm{log} \ g$, RV,
and $v\mathrm{sin}i$ are on average accurately recovered. This also suggests that there are no systematic trends between the recovered parameters and the input (i.e., true) values.
As expected, the scatter of the distributions increases with decreasing S/N and it provides an indication of the average uncertainty on the parameter estimates.
The results of the comparison between the recovered parameters and the input values provide a crucial benchmark for the procedure,
confirming that the method does not introduce any systematic effect.

It is important to note that the moderate spectral resolution of MUSE imposes a lower limit on the measurement of the rotational velocity. In particular, the analysis of the synthetic spectra shows that we are only poorly sensitive to detect $v\mathrm{sin}i$ values smaller than $\sim$ 50 km s$^{-1}$.
As a result, for stars in the slow rotation regime ($\lesssim$ 50 km s$^{-1}$) we can only provide upper limits. 

\section{Selection of the bona-fide sample}
\label{sec:membership}
We applied the fitting procedure described in section \ref{sec:spectral_fit} to all the extracted MUSE spectra of NGC 1783 to derive the best-fit values of $T_{\mathrm{eff}}$, $\mathrm{log} \ g$, RV and $v\mathrm{sin}i$ for each star. 

The following analysis is then applied only to a sub-sample of {\it bona-fide} stars selected as follows.
First, as shown in Fig. \ref{fig:cmd_snr}, our dataset samples a wide range of magnitudes and, as a consequence, of S/N, which ranges from 80 down to 2 for the fainter targets. 
Therefore, in order to ensure the highest reliability of our results, we performed the subsequent analysis 
only on stars with S/N (on the combined spectra) larger than $15$, which roughly corresponds to magnitude $m_{F438W} < 22.5$.

\begin{figure}[h!]
\centering
\includegraphics[width=0.47\textwidth]{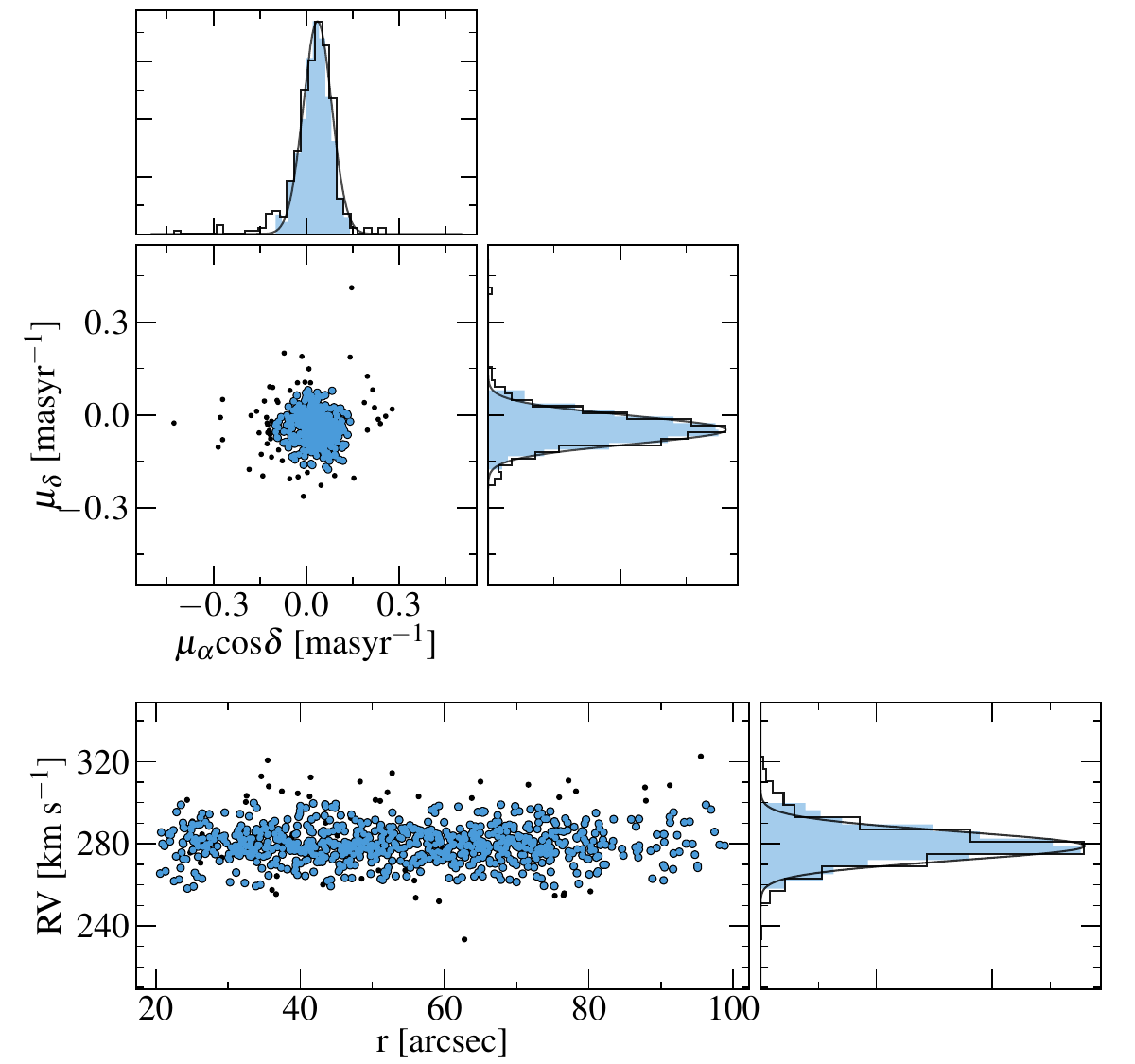}
\centering
\caption{\textit{Top panels}: VPD of NGC 1783 based on the PM measurements by \citet{Cadelano+22}. The blue circles show the MUSE targets selected as likely cluster’s member stars, while the black circles are the MUSE targets classified as non-members.
The smaller panels at the top and on the right show the distributions of both the PM components: the empty histograms represent the entire MUSE sample with the corresponding Gaussian fit indicated by the black curves, while the blue histograms are relative to the bona ﬁde cluster members.
\textit{Bottom panels}: the left panel shows the RVs of the MUSE sample with S/N$>15$ plotted as a function of the distance from the cluster center. 
The blue circles highlight the {\it bona-fide} sample selected according to 
the criteria discussed in the Section \ref{sec:membership}, while the black dots are the excluded targets.
The empty histogram on the left panel shows the number distribution of the entire RV sample, while the blue histogram corresponds to the final sub-sample of {\it bona-fide} stars.}
\label{fig:membre}
\end{figure}

Secondly, the following analysis (Section~\ref{sec:results}) targets likely cluster member stars. Cluster membership was defined by adopting a combination of proper motion (PM)-based and RV-based selections.
To do this, we used the PMs measured by \citet{Cadelano+22}. 
We performed a Gaussian ﬁt on both the PM components independently. Stars with PM smaller than $3\sigma_{\rm PM}$, where $\sigma_{\rm PM}$ is the best ﬁt Gaussian width, were selected as cluster’s members.
The vector-point diagram (VPD) of NGC 1783 and the distributions of the two PM components are shown in the upper panels of Fig. \ref{fig:membre}, with the stars selected as members highlighted in blue. 
We included also a selection using the RVs computed from the spectral fitting. 
The bottom left and right panels of Fig. \ref{fig:membre} show the RVs of the MUSE targets with S/N$>15$ as a function of the distance from the cluster center and the corresponding distribution, respectively. We first removed obvious outliers with RV$<230$ km s$^{-1}$ and RV$>327$ km s$^{-1}$, and then computed the $\sigma_{\rm RV}$ of the distribution through a Gaussian fit on the remaining stars. The targets having RV higher than 3$\sigma_{\rm RV}$ have been excluded.
For stars having the three velocity components, we adopted both the
PM-based and RV-based selection criteria, while for the targets with no PM measurements we used only the RVs to identify the members.
The combination of S/N and membership selections provide us with a {\it bona-fide} sample of 662 member stars, which are highlighted as blue circles in the bottom-left panel of Fig. \ref{fig:membre}. 

\section{Stellar rotational velocities}
\label{sec:results}
\setcounter{figure}{6}
\begin{figure*}[hb!]
\centering
\includegraphics[width=0.63\textwidth]{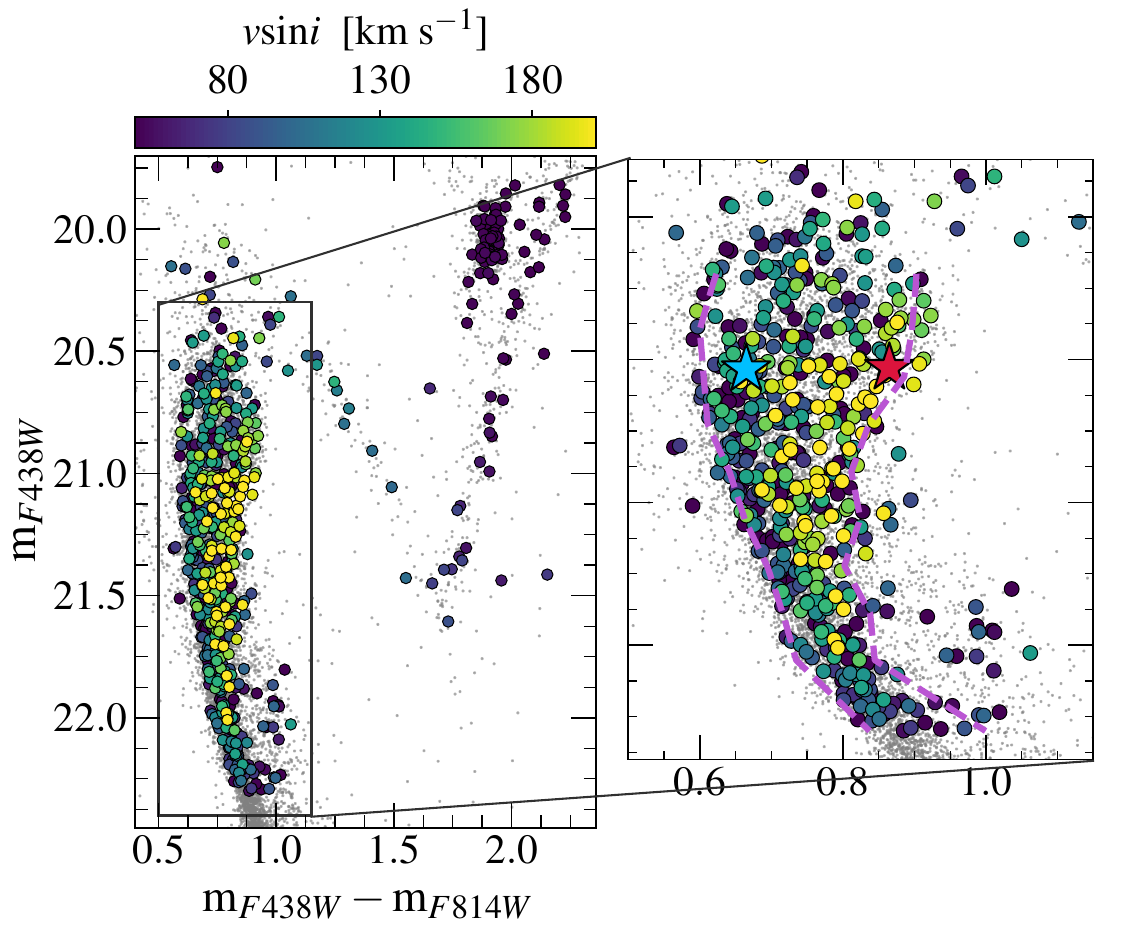}
\caption{CMD of NGC 1783 with the {\it bona-fide} stars color-coded according to their best-fit $v\mathrm{sin}i$ values. The gray box highlights the eMSTO region, which is shown as a zoomed-in view in the right panel.
The blue and red star symbols mark the slow-rotating target $\#44909$ and the fast-rotating target $\#43977$, respectively, for which the spectral fitting results are shown in Fig. \ref{fig:corner_new}, and their spectra are compared in Fig. \ref{fig:spectra}.
The violet dashed lines are the fiducial lines adopted to verticalize the color distribution.}
\label{fig:cmd_vsini}
\end{figure*}
\setcounter{figure}{5}
\begin{figure}[ht!]
\centering
\hspace{-0.3cm}
\includegraphics[width=0.48\textwidth]{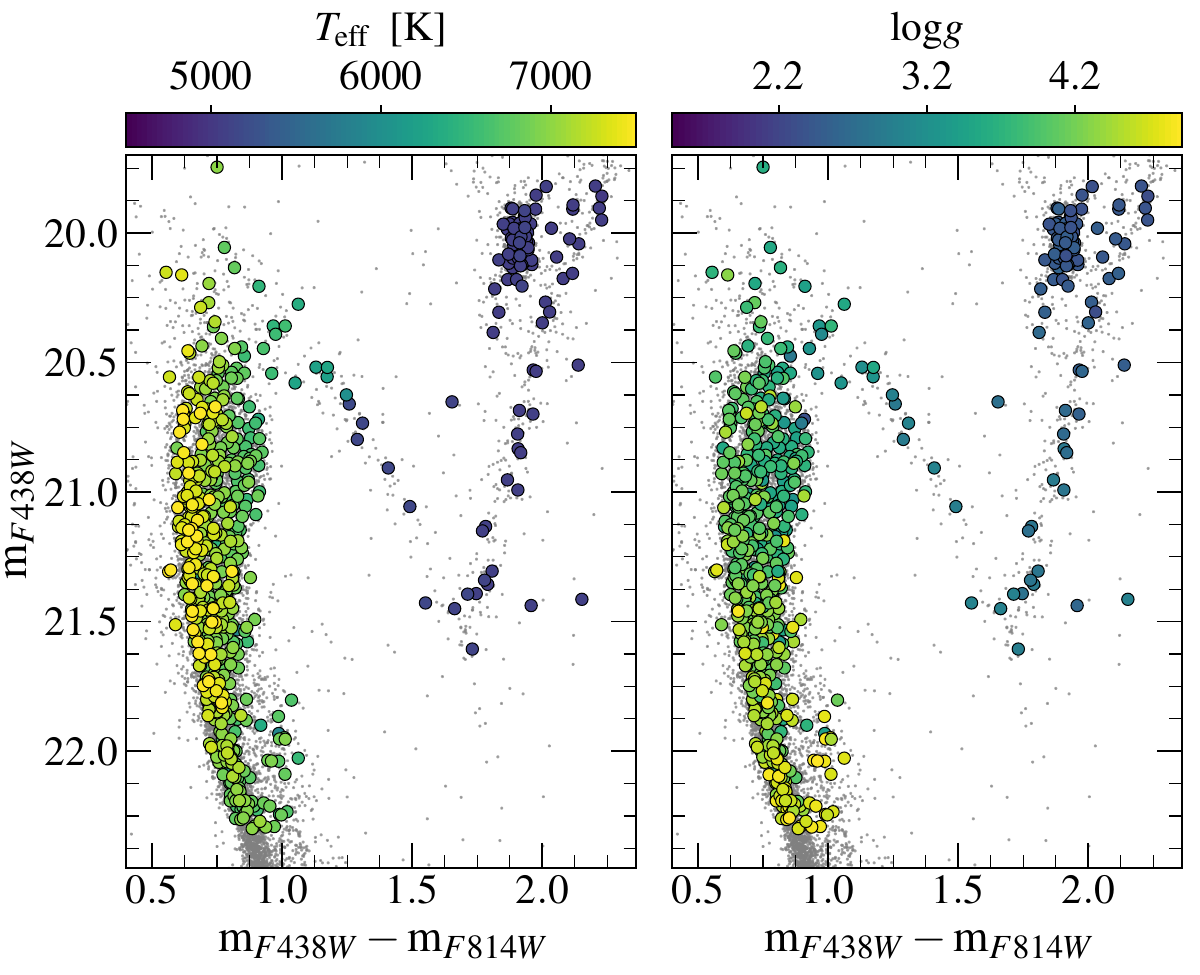}
\centering
\caption{CMDs of NGC 1783 with the stars of the {\it bona-fide} sample color-coded according to their values of $T_{\mathrm{eff}}$ (left panel) and $\mathrm{log} \ g$ (right panel).}
\label{fig:cmd_results}
\end{figure}
Fig. \ref{fig:cmd_results} shows the color-magnitude distribution of the {\it bona-fide} MUSE stars, color-coded by their best-fit values of $T_{\mathrm{eff}}$ and $\mathrm{log} \ g$ in the left and right panels, respectively.
As expected, we observe pretty clearly a correlation between $T_{\mathrm{eff}}$ and the photometric color, and between $\mathrm{log} \ g$ and stellar luminosity. This represents an independent test about the reliability of the results obtained with our fitting method.
In Figure~\ref{fig:cmd_vsini}, the {\it bona-fide} stars are color-coded by their best-fit value of $v\mathrm{sin}i$. 
The figure shows that the rotational velocities cover a wide range of values, going from $v\mathrm{sin}i < 50$ km s$^{-1}$
(consistent with no or slow rotation) to $v\mathrm{sin}i\sim 250$ km s$^{-1}$ (fast rotation).
In particular, signatures of weak rotation are found among sub-giant stars while red giants are non-rotating or at most slow rotators. 
Stars along the eMSTO sample the full range of the observed $v\mathrm{sin}i$ values, as shown in the zoomed-in view of Fig.~\ref{fig:cmd_vsini}.

The blue and red star symbols in the right panel of Fig. \ref{fig:cmd_vsini} are two representative cases of slow and fast rotator stars, for which the results of the fitting method are shown in the left and right panels of Fig. \ref{fig:corner_new}, respectively.
The two targets were selected on the opposite edge of the eMSTO, i.e. with different photometric colors, but with approximately the same magnitude (m$_{F438W}\sim$21) and about the same S/N.
In the bottom panels of Fig. \ref{fig:corner_new}, the best-ﬁt synthetic spectra (red and blue curves) are overplotted to the observed ones (black curve), showing an excellent match. 
The one- and two-dimensional posterior probability distributions for all the parameters considered in the analysis are represented as corner plots in the top panels.
The best-fit analysis (the results are given in the top panels of Fig. \ref{fig:corner_new}) yields a temperature difference of $\sim$ 350 K between the two targets, with the rapidly rotating star being cooler. We also find that there is a difference of $\sim$ 170 km s$^{-1}$ in rotational velocities between the two stars. 
The effect of stellar rotation on the spectral profile is clearly visible in the bottom panels of Fig. \ref{fig:corner_new}.
The direct comparison of the two spectra in Fig. \ref{fig:spectra} shows even more clearly how the rapid rotation significantly broadens the spectral lines. \\
We note that the posterior probability distribution of $v\mathrm{sin}i$ exhibits a well-defined peak for the fast rotator star, while for the slow rotator, it shows a roughly constant plateau for values smaller than $\sim 40$ km s$^{-1}$. As already pointed out in Section \ref{sec:spectral_fit}, this is due to the intrinsic limited ability 
to measure small $v\mathrm{sin}i$ values as a result of the relatively poor MUSE spectral resolution.\\
\setcounter{figure}{7}
\begin{figure*}[!]
  \centering
  \begin{minipage}[]{0.45\textwidth}
        \includegraphics[width=0.94\textwidth]{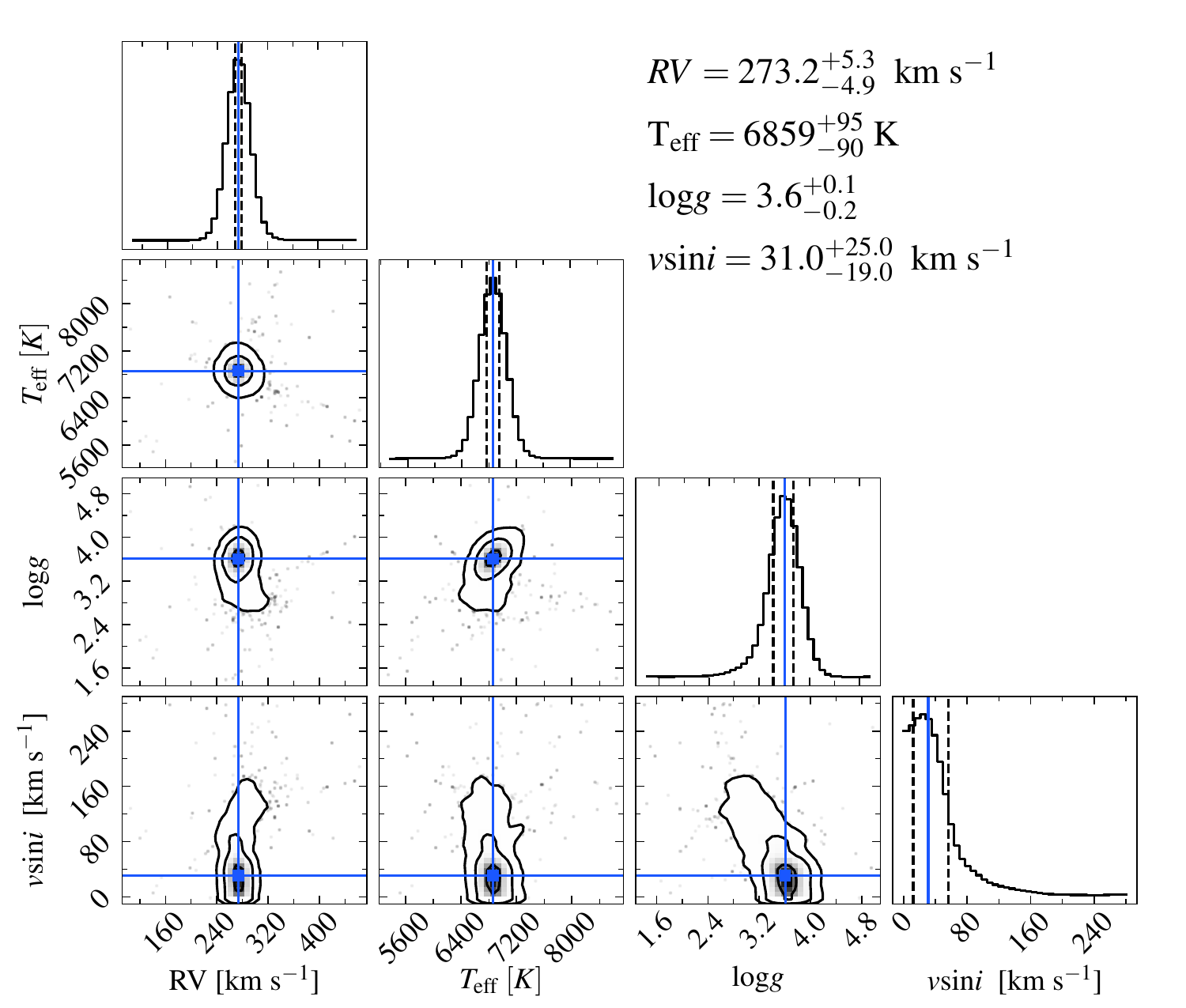}\vspace{0.10cm}
        \includegraphics[width=0.94\textwidth]{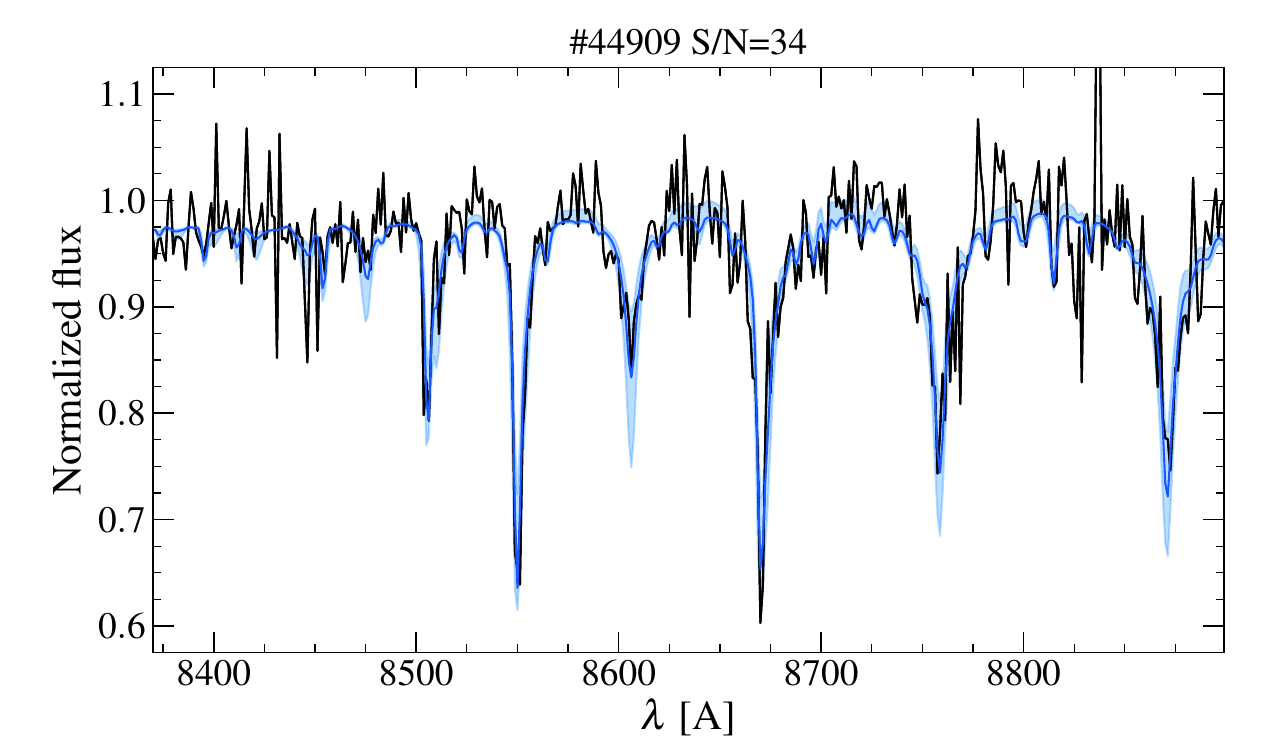}
  \end{minipage}
  \hspace{0.4cm}
  \begin{minipage}[]{0.45\textwidth}
          \includegraphics[width=0.94\textwidth]{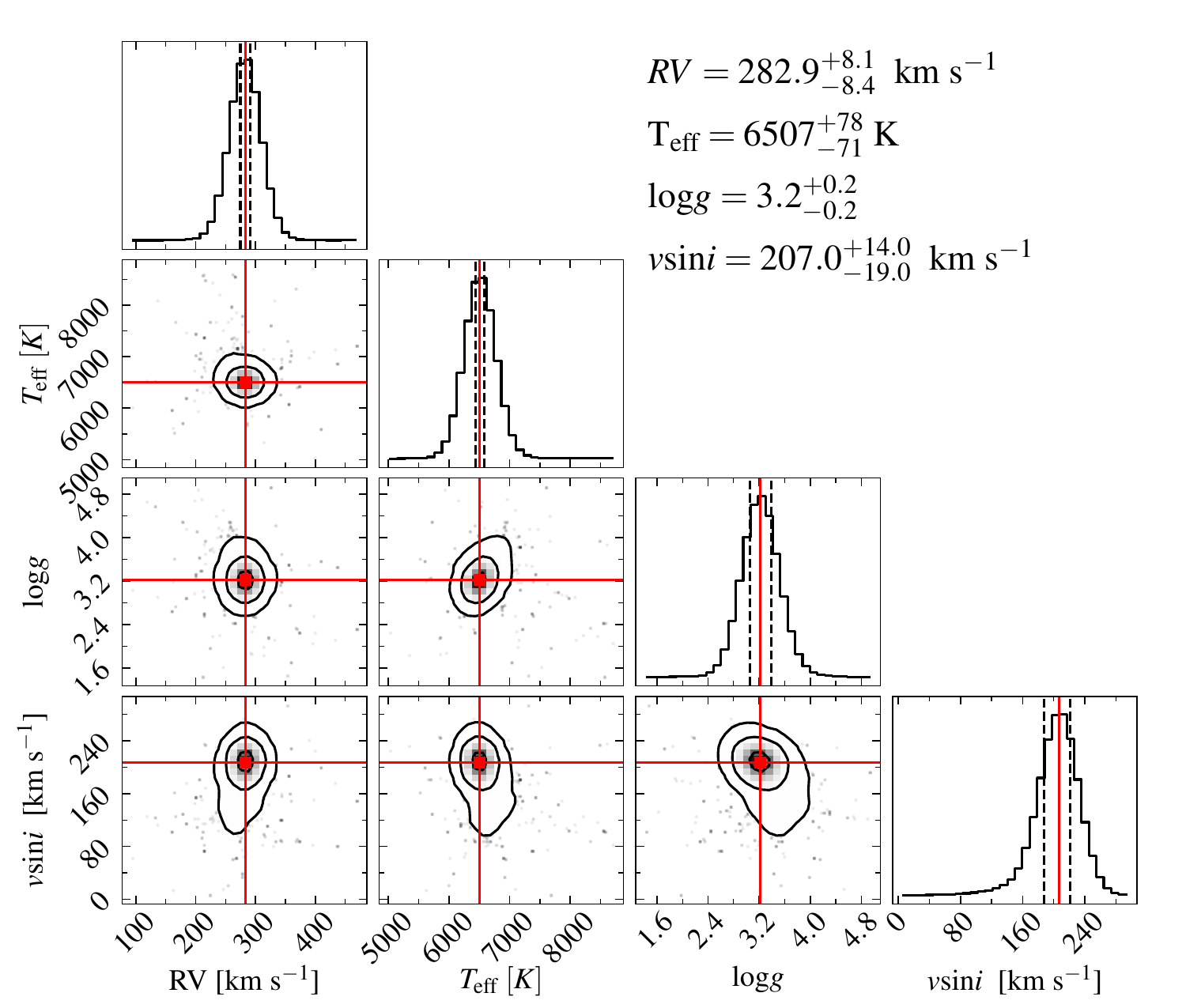}\vspace{0.10cm}
          \includegraphics[width=0.94\textwidth]{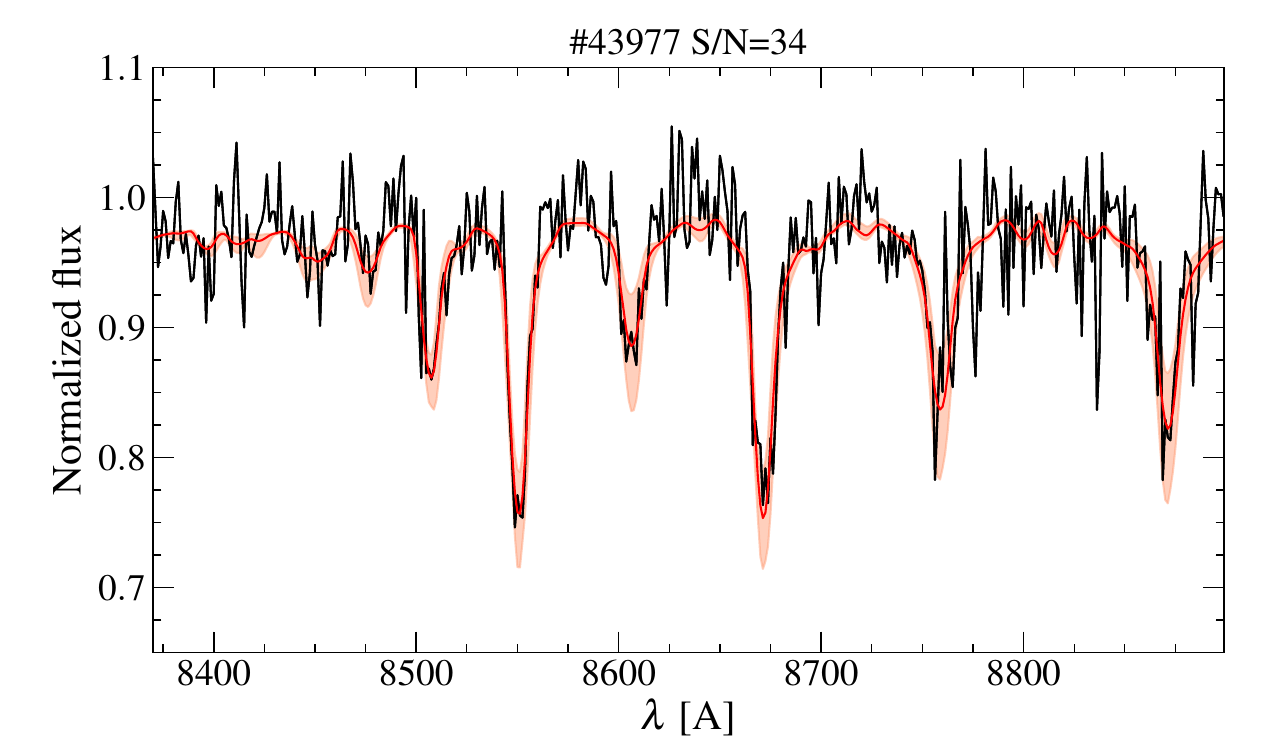}
  \end{minipage}
        \caption{Results of the spectral fitting procedure for the target $\#44909$ on the {\it left panels} and the target $\#43977$ on the {\it right panels}.  
\textit{Top panels}: corner plots showing the one- and two-dimensional posterior probability distributions for all the parameters derived from the MCMC algorithm. The contours correspond to the 68\%, 95\%, and 99\% confidence levels, and the best-fit parameter values are also labeled.
\textit{Bottom panels}: the black curves are the observed MUSE spectra in the Ca II wavelength region, plotted together with the corresponding best-fit synthetic models shown by the blue and red solid curves for the targets $\#44909$ and $\#43977$, respectively. The shaded blue and red regions represent the solutions within the $1\sigma$ confidence level.}
        \label{fig:corner_new}
\end{figure*}
Figure \ref{fig:hist1} shows the distribution of rotational velocities of the eMSTO stars. The figure highlights the wide range of $v\mathrm{sin}i$ values sampled by the eMSTO targets, spanning from slow/no rotation to fast rotation. The distribution shows that there is a non negligible fraction 
($\sim 18\%$) of stars having slow rotation velocity ($v\mathrm{sin}i < 50$ km s$^{-1}$). At larger rotational velocities, the number of stars decreases progressively as
$v\mathrm{sin}i$ increases. About $75\%$ of the stars have $v\mathrm{sin}i$ values between $50$ and $200$ km s$^{-1}$. The distribution also shows a tail at higher rotational velocities, with $\sim 7\%$ of stars exceeding $v\mathrm{sin}i > 200$ km s$^{-1}$.
Additionally, in Fig. \ref{fig:cmd_vsini}, we observe that the majority of the fastest rotating stars ($v\mathrm{sin}i > 200$ km s$^{-1}$) are concentrated in the magnitude range $21 < $ m$_{F438W} < 21.3$, whereas in the brightest part of the eMSTO, fast rotators exhibit velocities below 200 km s$^{-1}$ (see also Fig. \ref{fig:vsini_bin.mag_hist}). This behavior is also observed in NGC 1846 by \citet{Kamann+20} and may suggest that the brighter eMSTO stars have been braked during the end of their main sequence lifetimes. 
However, the evolution of rotational velocity during the main sequence is complex and depends on multiple factors, including stellar expansion, the efficiency of angular momentum transport, and wind mass loss 
\citep[see, e.g.,][]{Ekstrom+08, Zorec+12, Hastings+20}.
Consequently, drawing a definitive interpretation of the observed trend remains challenging, but future theoretical models may provide valuable insights to better understand these observations. On the other hand, at fainter magnitudes (m$_{F438W} > 22$) the number of fast rotators decreases drastically likely due to the effect of magnetic braking. 
Interestingly, the zoomed view around the eMSTO region in Fig. \ref{fig:cmd_vsini} shows clearly that there is a correlation between the derived $v\mathrm{sin}i$ values and the color of the eMSTO stars, with $v\mathrm{sin}i$ increasing as the color increases.
This trend is in very nice agreement with results found in previous studies of young and massive clusters in the Magellanic Clouds and Galactic young open clusters \citep[see e.g.][]{Bastian+18, Kamann+18b, Kamann+20, Cordoni+21, Cordoni+24}, further 
strengthening the role played by stellar rotation in shaping the CMDs of young and intermediate age clusters.

\begin{figure}[ht!]
\centering
\includegraphics[width=0.43\textwidth]{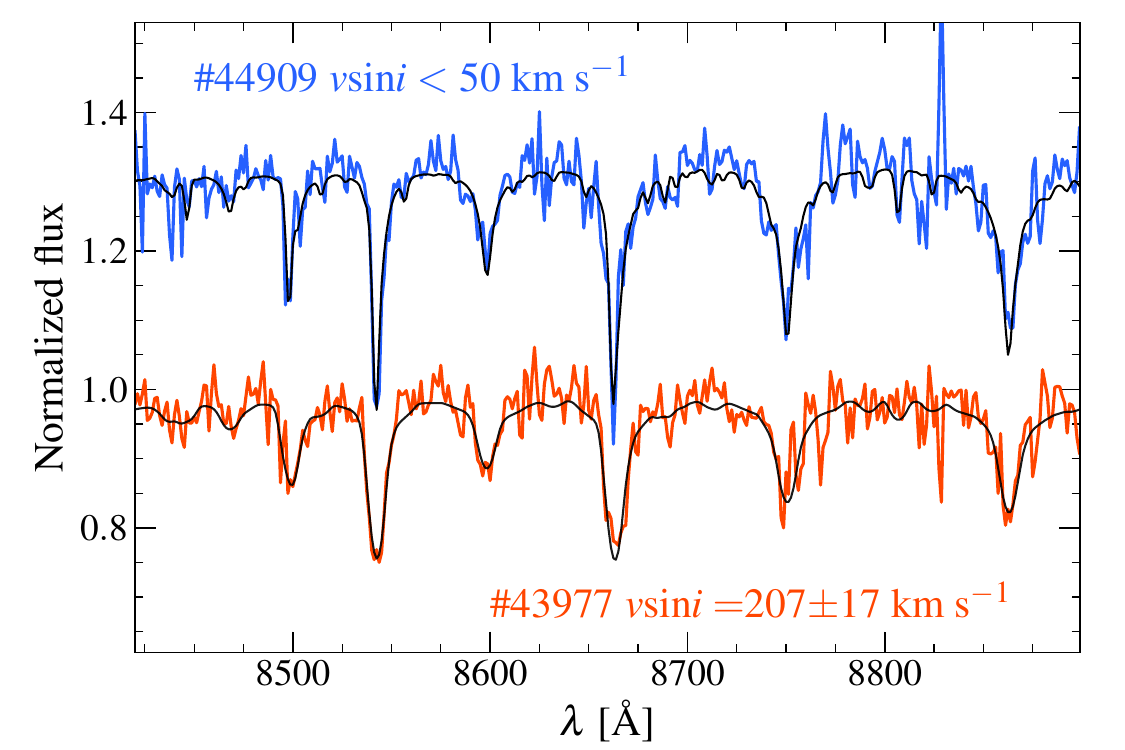}
\caption{Comparison of MUSE spectra in the Ca II region of two eMSTO stars with a low value of $v\mathrm{sin}i$ (target $\#44909$, blue curve), consistent with no/slow rotation, and a high rotational velocity (target $\#43977$, red curve). 
For both spectra, the black solid curves show the respective best-fit synthetic models. Note that the blue and red spectra are the same as those shown in Fig. \ref{fig:corner_new}, and correspond to the targets marked by the star symbols in Fig. \ref{fig:cmd_vsini}, using the same color code.}
\label{fig:spectra}
\end{figure}
\setcounter{figure}{10}
\begin{figure*}[hb!]
\centering
\includegraphics[width=0.67\textwidth]{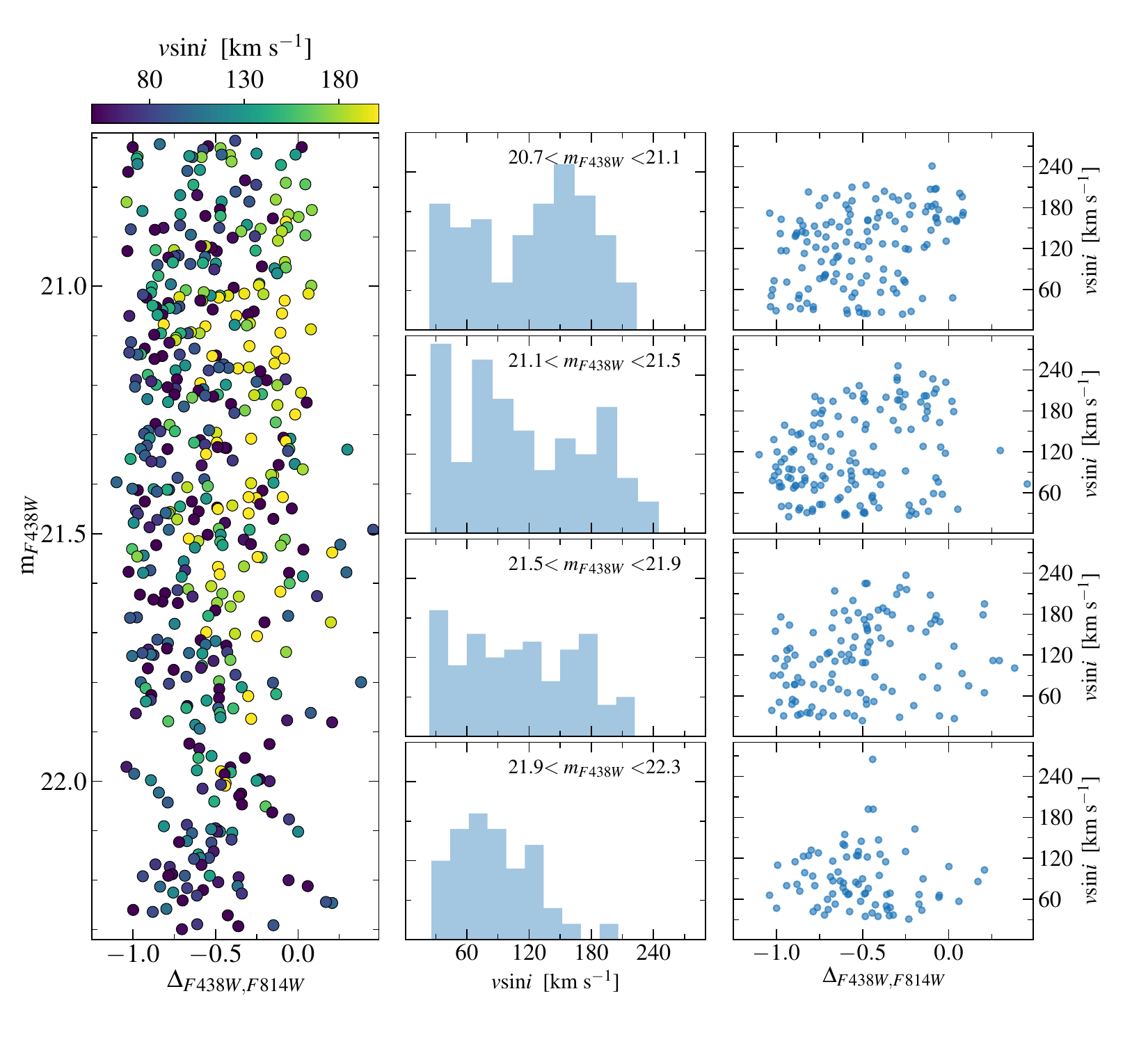}
\centering
\caption{The left panel displays the verticalized color distribution of the eMSTO of NGC 1783, with the targets color-coded according to the their $v\mathrm{sin}i$ values.
The central panels show the histograms of the $v\sin i$ measurements for the targets in the magnitude bins labeled at the top of each panel.
The right panels represent the $v\sin i$ values as a function of the pseudo-color $\Delta_{F438,F814W}$ of the targets in the same magnitude bins as in the central panels.}
\label{fig:vsini_bin.mag_hist}
\end{figure*}
\setcounter{figure}{9}
\begin{figure}[ht!]
\centering
\includegraphics[width=0.39\textwidth]{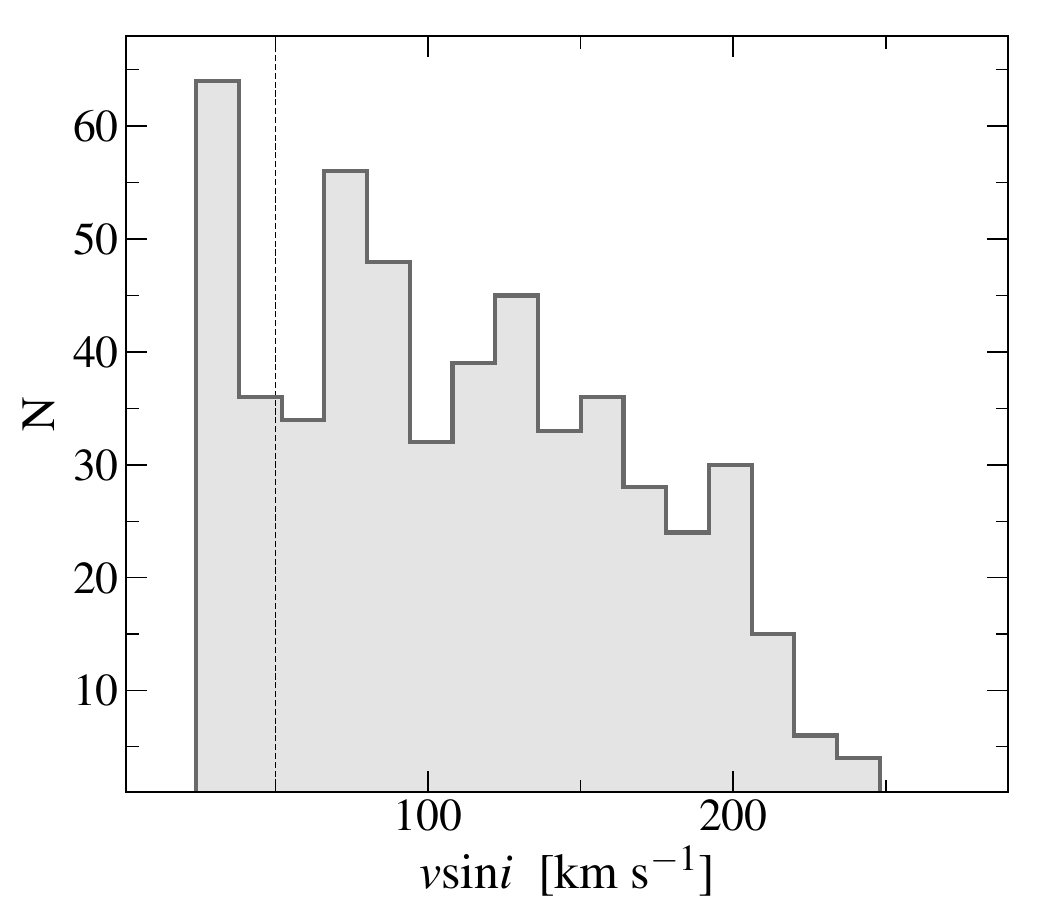}
\centering
\caption{Distribution of $v\mathrm{sin}i$ of the eMSTO stars. Note that all measurements below the threshold value of 50 km s$^{-1}$ (dashed line) represent upper limits (see Section~4 for details). 
}
\label{fig:hist1}
\end{figure}
\setcounter{figure}{11}
\begin{figure*}[h!]
\centering
\includegraphics[width=0.8\textwidth]{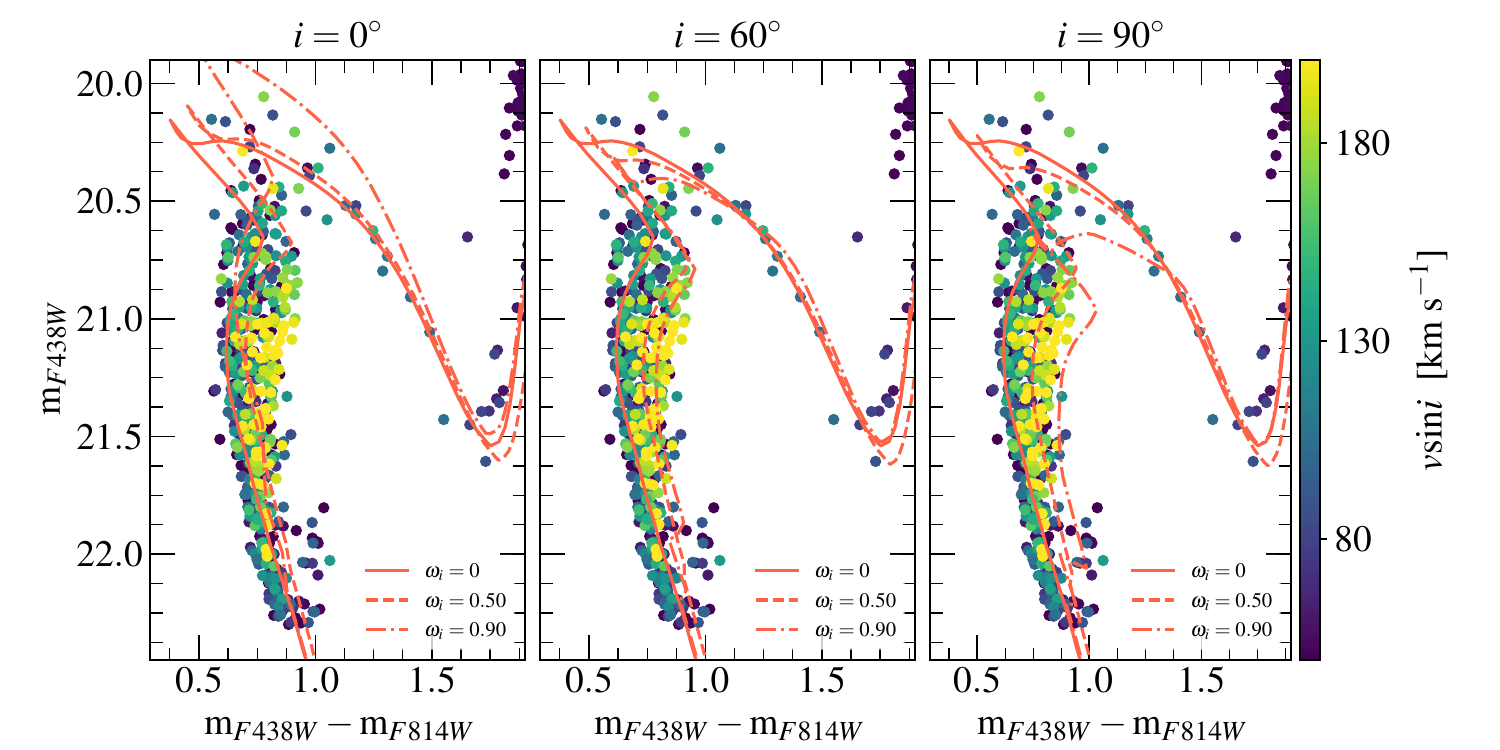}
\centering
\caption{Comparison between the CMD of the {\it bona-fide} stars in the eMSTO region of NGC 1783 and a set of PARSEC rotating isochrones (red curves) computed for an age of 1.5 Gyr, a metallicity of [Fe/H]$= -0.35$ and different initial rotation rates ($\omega_{i} = 0.0, 0.5, 0.9$). In each panel the same set of isochrones is shown while varying the inclination angles of the rotation axis with respect to the line of sight, $i = 0\degr$ (left), $i = 60\degr$ (central), and $i = 90\degr$ (right). The targets are color-coded according to their $v\mathrm{sin}i$ values.}
\label{fig:cmd_models}          
\end{figure*}

To further investigate the behavior of the $v\mathrm{sin}i$ along the eMSTO and the upper MS, we followed the same approach used in \citet[][see also \citealt{Bastian+18}]{Kamann+20}.
We selected the eMSTO and MS stars in the magnitude range 20.7 $\leq \mathrm{m}_{F438W} \leq $ 22.3 and limiting the color range to 
0.59 $\leq \mathrm{m}_{F438W} - \mathrm{m}_{F814W} \leq $ 1.0 to remove any outliers. Using this sample of stars, we computed the 4th and 96th percentiles of the distribution in $\mathrm{m}_{F438W} - \mathrm{m}_{F814W}$ colour as a function of $\mathrm{m}_{F438W}$ magnitude, defining two fiducial lines at the bluest and reddest edges of the MS, respectively.
The resulting MS edges are shown as dashed lines in Fig. \ref{fig:cmd_vsini}.
We then measured the normalized color distance of each star to the red edge, defining the pseudo-color $\Delta_{F438W,F814W}$.
The resulting verticalized color distribution is shown in the left panel of Fig. \ref{fig:vsini_bin.mag_hist}.
The central panels of the same figure display the $v\mathrm{sin}i$ distributions by splitting the sample into four magnitude bins, while the right panels show the $v\mathrm{sin}i$ values as a function of pseudo-color for the same magnitude bins as in the central panels.
The panels are shown in order of increasing magnitude, and the magnitude range of each bin is labeled at the top of each central panel.\\
The histograms in the central panels show that while the three brightest bins sample the full range of rotational velocities, ranging from slow (or no rotation) to fast rotation, the faintest bin lacks almost entirely of stars with $v\mathrm{sin}i>140$ km s$^{-1}$. The lack of a high-rotating component suggests that magnetic braking process start becoming efficient in slowing down stars in the mass regime sampled in the faintest magnitude bin ($\sim1.2$  M$_{\odot}$). Interestingly,
there is also a hint of bimodality in the $v\mathrm{sin}i$ distribution in the two brighter magnitude bins. 
Finally, in agreement with what observed in Figure\ref{fig:cmd_vsini}, 
the three brighter bins of the right panels of Fig. \ref{fig:vsini_bin.mag_hist} indicate a mild correlation between the rotational velocity and the pseudo-color of the targets, in the sense of increasing $v\mathrm{sin}i$ values toward redder colors, as 
already observed in previous works \citep[see e.g.,][]{Bastian+18, Kamann+20}.
An ideal correlation between stellar color and rotational velocity is likely attenuated by the effect of multiple factors, including different inclination angles of the rotation axis and the presence of unresolved binaries.

To better illustrate the effect of stellar rotation in shaping the eMSTO, in Fig. \ref{fig:cmd_models} we compare the observed CMD in the eMSTO region with a set of PARSEC\footnote{\url{https://stev.oapd.inaf.it/PARSEC/index.html}} v2.0 isochrones \citep{Nguyen+22}
with different initial angular rotation rates ($\omega_i$) and with different inclination angles of the line of sight 
with respect to the stellar rotation axes, $i$.
The rotation rate is defined as $\omega = \Omega / \Omega_c$, where $\Omega$ is the angular velocity, and $\Omega_c$ is the critical angular velocity (or breakup velocity), i.e. the angular velocity at which the centrifugal force is equal to the effective gravity at the equator.
In particular in Fig. \ref{fig:cmd_models} we show three sets of isochrones computed assuming an age of 1.5 Gyr, [Fe/H]$=-0.35$ (as appropriate for NGC~1783) and different initial rotation rates, $\omega_{i} = 0.0, 0.5, 0.9$. Each panel shows the same set of isochrones but for different inclination angles, $i = 0 \degr$ (left), $i =60 \degr$ (middle) and $i =90 \degr$ (right).
To overplot the isochrones to the CMD, we used a distance modulus of $(m - M)_0 = 18.47$ and an extinction value of $E(B - V) = 0.06$\footnote{
These values of distance modulus and extinction differ slightly from those adopted by \citet{Cadelano+22} for the same photometric dataset. However, this difference is not of concern, as it is due to the use of different isochrone models in the two studies.}.
For all inclination angles, the observed trend between eMSTO stars and $v\mathrm{sin}i$ is nicely matched (at least qualitatively) by the models.
Interestingly, the trend shown in Fig. \ref{fig:cmd_models} is consistent with other sets of models including the effect of rotation, such as the MIST models \citep{Choi+16},
while it goes in the opposite direction than what predicted by the SYCLIST models \citep[see,][]{Brandt_2015, Eggenberger+21}, in which fast-rotating models appear bluer than the slow-rotating ones in the TO region. 
Interestingly we note that at odds with the PARSEC and MIST models, the SYCLIST ones adopt efficient rotational mixing.  
Hence, the qualitative agreement between the observed distribution of $v\mathrm{sin}i$ and predictions
by the PARSEC (and MIST) models in the specific case of NGC~1783, together with previous analysis \citep[e.g.,][]{marino+18b,Kamann+20,kamann+23}, would possibly suggest that rotational mixing is inefficient.

The isochrones in Fig. \ref{fig:cmd_models} also show the effects of different inclination angles on the color-magnitude distribution of stars along the eMSTO. 
The rotating isochrones with an inclination angle of $i = 0 \degr$ (pole-on, left panel) appear brighter and bluer than the isochrones with the same rotation rate but with $i = 90 \degr$ (equator-on, left panel). As a result, relative differences among models with different rotation rates increase for increasing values of $i$ thus showing that also projection effects contribute in shaping the color distribution of rotating stars along the eMSTO.

\setcounter{figure}{13}
\begin{figure*}[hb!]
\centering
\includegraphics[width=0.85\textwidth]{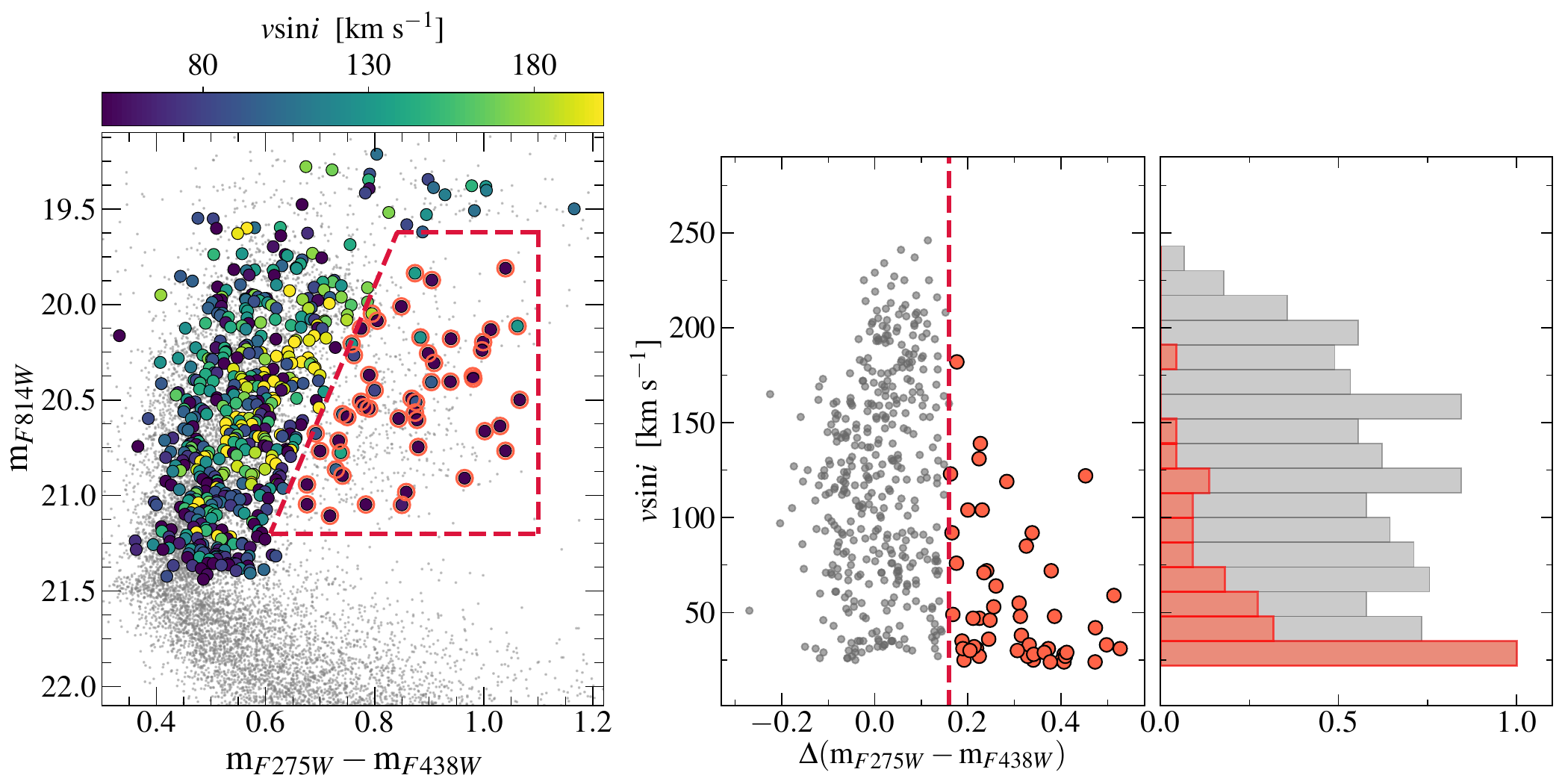}
\centering
\caption{Left panel: CMD focusing on the eMSTO region of NGC 1783. The MUSE targets are represented as large circles, color-coded according to their $v\sin i$ values. 
The red dashed box outlines the selection region for UV-dim stars (see text), which are highlighted by open red circles.
Central panel: $v\sin i$ of the eMSTO targets in the magnitude range $19.6< m_{F814W} < 21.2$ as a function of the verticalized color. The dashed red line marks the selection edge for the UV-dim stars, corresponding to $\Delta$($m_{\rm F275W}-m_{\rm F438}$) $=0.16$, with the selected UV-dim stars indicated by red circles. 
Right panel: normalized histograms of the $v\sin i$ distributions for the stars shown in the central panel. The gray histogram represents the eMSTO stars, while the red histogram shows the $v\sin i$ distributions for the sub-sample of UV-dim stars.}
\label{fig:plots_uvdim}
\end{figure*}
\setcounter{figure}{12}
\begin{figure}[ht!]
\centering
\hspace{-0.7cm}
\includegraphics[width=0.5\textwidth]{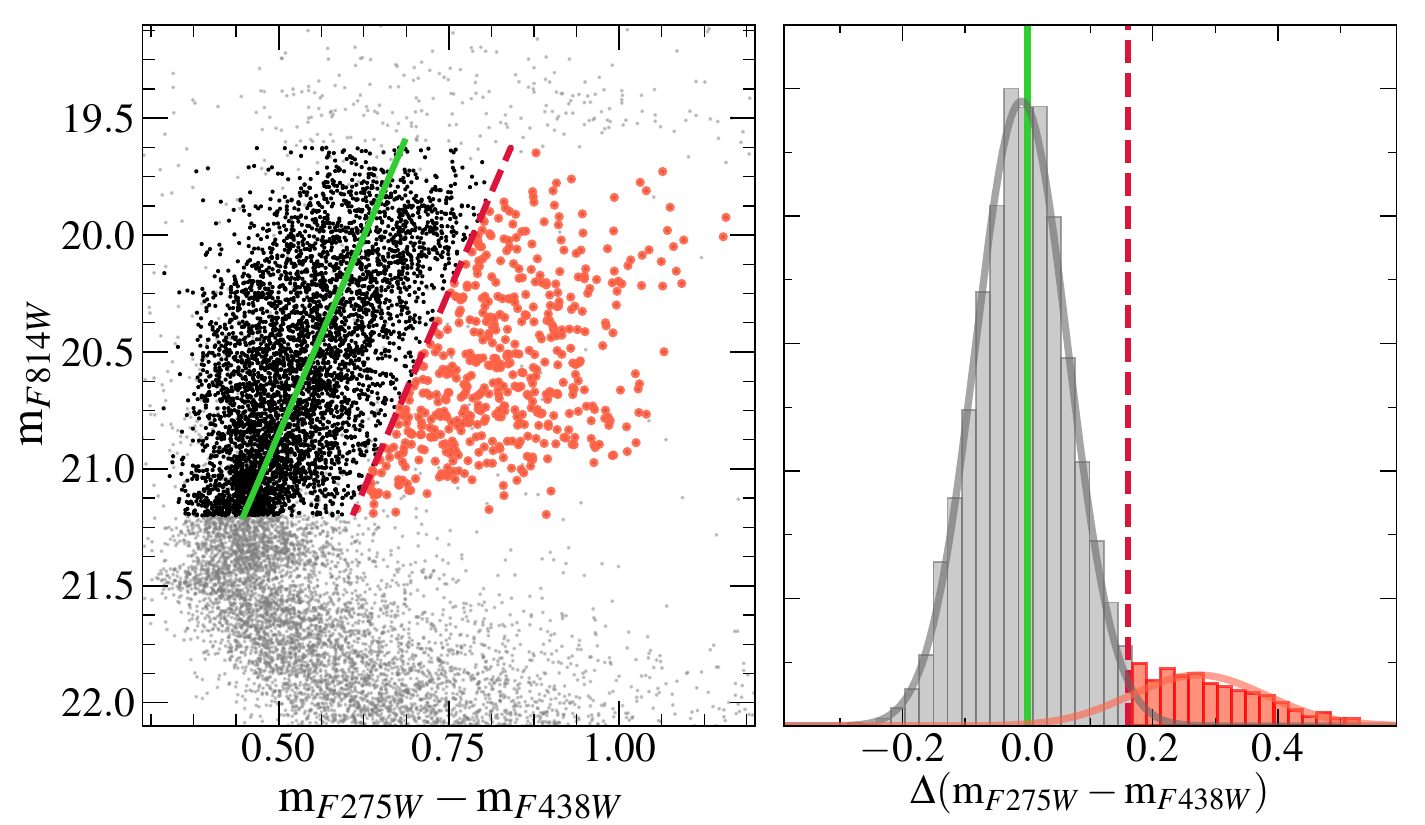}
\centering
\caption{Left panel: $m_{\rm F275W}-m_{\rm F438W}$, $m_{\rm F814W}$ CMD of NGC 1783. The black circles show the selected MSTO stars, while the red circles represent the stars identified as UV-dim. The green solid and the red dashed lines indicate the MSTO ridge line and the selection threshold for the UV-dim stars (see right panel), respectively.
Right panel: histogram of the verticalized color distribution 
($\Delta$($m_{\rm F275W}-m_{\rm F438}$)) for the MSTO stars. 
The red dashed line represents the threshold in the verticalized color, $\Delta$($m_{\rm F275W}-m_{\rm F438}$) $=0.16$, used to select the UV-dim stars, which are highlighted in red. Gray and red Gaussian curves indicate the respective GMM fit components. The green line is the same as in the left panel.}
\label{fig:sele_uvdim}
\end{figure}

\section{UV-dim stars}
\label{sec:uvdim}
NGC 1783 is well-known to host a large population of UV-dim stars and it has been the subject of numerous investigations  in this respect \citep{milone+23a,d'antona+23,martocchia+23,milone+23b}. Various hypotheses have been proposed regarding the nature of the UV-dim stars, however a definitive understanding remains elusive.
In this work, we used MUSE spectra to add new constraints on the stellar rotational properties of these peculiar objects. 

\subsection{Photometric selection of UV-dim stars}
To ensure a sufficiently high statistic, UV-dim stars were selected from the 
HST photometric catalog \citep{Cadelano+22} by schematically following the procedure described in \citet{martocchia+23}. In particular, we first selected eMSTO stars in the ($m_{\rm F438W}-m_{\rm F814W}$, $m_{\rm F814W}$) CMD and in the magnitude range $19.6< m_{F814W} < 21.2$. Then, we computed the mean ridge line of the thus selected eMSTO stars
in the ($m_{\rm F275W}-m_{\rm F438W}$, $m_{\rm F814W}$) UV CMD.
To this aim we computed the mean of a smoothing spline fit of a first-degree polynomial (green solid line in Fig. \ref{fig:sele_uvdim}) to the color distribution of stars. Finally, the distance of each eMSTO star from the ridge line was computed, obtaining the $\Delta$($m_{\rm F275W}-m_{\rm F438}$) verticalized color.  
The distribution of the verticalized color is well-fitted by two Gaussian functions, as shown in the right panel of Fig. \ref{fig:sele_uvdim}. The fit was performed by using a Gaussian mixture model (GMM), and the Bayesian Information Criterion (BIC) was employed to determine the optimal number of components.
To identify the UV-dim stars, we set a color threshold at $\Delta(m_{\rm F275W}-m_{\rm F438W}) = 0.16$ (red dashed line in Fig. \ref{fig:sele_uvdim}). This value corresponds approximately to the intersection point of the two Gaussian components. 
Specifically, stars with $\Delta$($m_{\rm F275W}-m_{\rm F438}$)$>0.16$ were selected as UV-dim stars and are highlighted with red circles in the CMD shown in the left panel of Fig. \ref{fig:sele_uvdim}.

The left panel of Fig. \ref{fig:plots_uvdim} shows the ($m_{\rm F275W}-m_{\rm F438W}$, $m_{\rm F814W}$) CMD of the {\it bona-fide} MUSE targets color-coded according to their $v\mathrm{sin}i$ values. The dashed red box highlights the region occupied by the UV-dim stars, which are marked with open red circles.
The oblique edge of the box represents the adopted color threshold ($\Delta$($m_{\rm F275W}-m_{\rm F438}$)$=0.16$) while the two horizontal lines correspond to the previously defined magnitude limits ($m_{\rm F814W} = 19.6$ and $m_{\rm F814W} = 21.2$).

\subsection{Properties of the UV-dim stars}
A preliminary visual inspection of Fig. \ref{fig:plots_uvdim} suggests that the majority of UV-dim stars are slow rotators. This trend becomes even more evident when examining the right panels of Fig. \ref{fig:plots_uvdim}. The central panel shows the $v\mathrm{sin}i$ of the eMSTO stars (in the magnitude range $19.6< m_{F814W} < 21.2$, gray circles) as a function of the verticalized color. The UV-dim stars, as selected in the left panel, are indicated by red circles, and the dashed red line marks the color threshold $\Delta$($m_{\rm F275W}-m_{\rm F438}$)$=0.16$. The right-hand panel presents the normalized histograms of the $v\mathrm{sin}i$ distribution for both the entire eMSTO sample (in gray) and the UV-dim stars (in red). 
The distribution clearly shows that the majority ($\sim60\%$) of UV-dim stars are non-rotating or at most slow rotators with $v\mathrm{sin}i<50$ km s$^{-1}$. 
The remaining 40\% is distributed along a declining tail for increasing values of $v\mathrm{sin}i$ almost ending at $\sim 150$ km s$^{-1}$ and with one single star with $\sim 180$ km s$^{-1}$, which is however very close to the blue edge of our selection, meaning that its classification as a UV-dim star may be uncertain.

The observational evidence that the UV-dim stars in NGC 1783 are preferentially slow-rotators appears to be in tension with the results of the spectroscopic analysis based on MUSE data performed by \citet{kamann+23} for the very young massive cluster NGC~1850, and with the results presented by \citet{martocchia+23} based on photometric arguments.  
In NGC~1850 UV-dim stars are among the fastest rotators in the cluster, with rotational velocities reaching values of up to $\sim250$ km s$^{-1}$. Interestingly, most of the UV-dim stars in NGC 1850 show also clear evidence of disk-like features in their spectra (such as a number of narrow Fe II and Si II absorption lines), which  suggest they are likely fast-rotating Be stars that are seen nearly equator-on (shell stars). 
To further explore this scenario also in the case of NGC~1783, we created two stacked spectra: one combining all the selected UV-dim stars, and the other including eMSTO stars with $v\sin i < 80$ km s$^{-1}$.
This approach enables a direct comparison in terms of stellar rotation.
We first checked for the presence of the absorption feature characteristic of shell stars \citep[see][]{kamann+23} in the UV-dim stacked spectrum. 
The top and the middle panels of Fig. \ref{fig:stacked_spectra} show that the Si II and Fe II lines of the UV-dim stacked spectrum (red curve) are slightly stronger than those observed in the spectrum obtained for eMSTOs (blue curve), thus possibly suggesting that UV-dim stars may have a disk also in the case of NGC 1783. While we note that the differences observed here are significantly smaller than those observed in the case of NGC~1850, we also stress that MSTO stars in NGC~1783 are significantly cooler ($\sim7000$ K) than in NGC~1850 ($\sim12000$ K) and that the strength of disk-related lines depends on the properties (e.g., mass and density) of the disks. \\
\setcounter{figure}{14}
\begin{figure}[h!]
\centering
\includegraphics[width=0.35\textwidth]{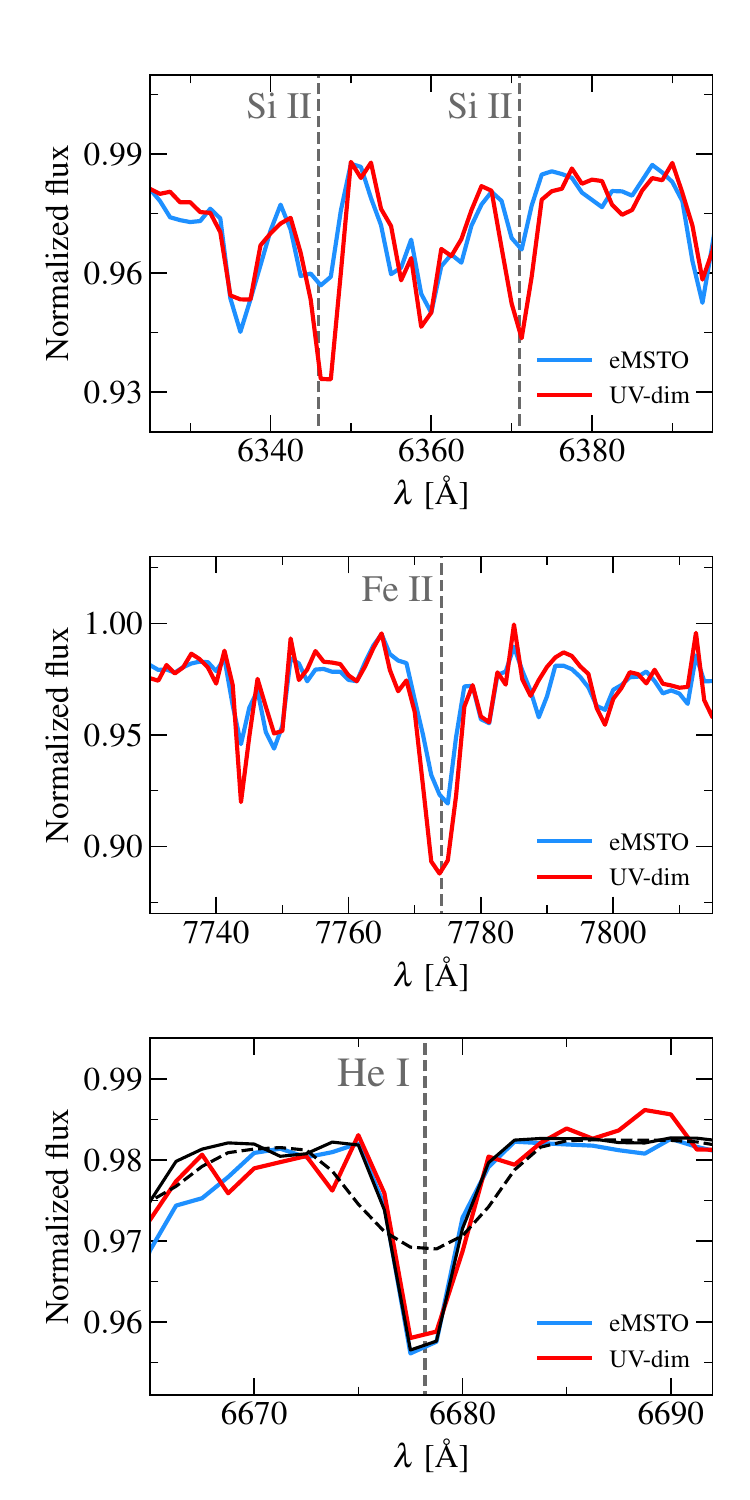}
\centering
\caption{Comparison between the stacked spectra obtained by combining the UV-dim stars and slow-rotating eMSTO stars, shown by the red and blue curves, respectively. In the bottom panel, the solid and dashed black curves represent a synthetic spectrum computed with $v\sin i = 60 $ km s$^{-1}$ and $v\sin i = 200 $ km s$^{-1}$, respectively.}
\label{fig:stacked_spectra}
\end{figure}
In the hypothesis that also the UV-dim stars in NGC~1783 are shell stars, then the rotational velocity constrained by using the Balmer and Paschen lines may be affected by the presence of the disk.
In particular, the presence of the disk can lead to narrow cores of the hydrogen lines, resulting in smaller $v\sin i$ values.
As an independent check on the derived rotational velocities, we used the HeI line at 6678.2 $\AA$ of the UV-dim stars stacked spectrum, which is the only non-hydrogen features not affected by the shell. We find that, in agreement with what obtained by using the fitting procedure described in Section \ref{sec:spectral_fit}, the HeI width is compatible with a small rotation of $v\sin i < 70$ km s$^{-1}$ (see bottom panel in Fig. \ref{fig:stacked_spectra}). Interestingly, a similar result is found when using the stacked spectrum of the slow-rotator eMSTOs, as shown by the blue curve in the bottom panel of Fig. \ref{fig:stacked_spectra}.
\begin{figure*}[ht!]
\centering
\includegraphics[width=0.97\textwidth]{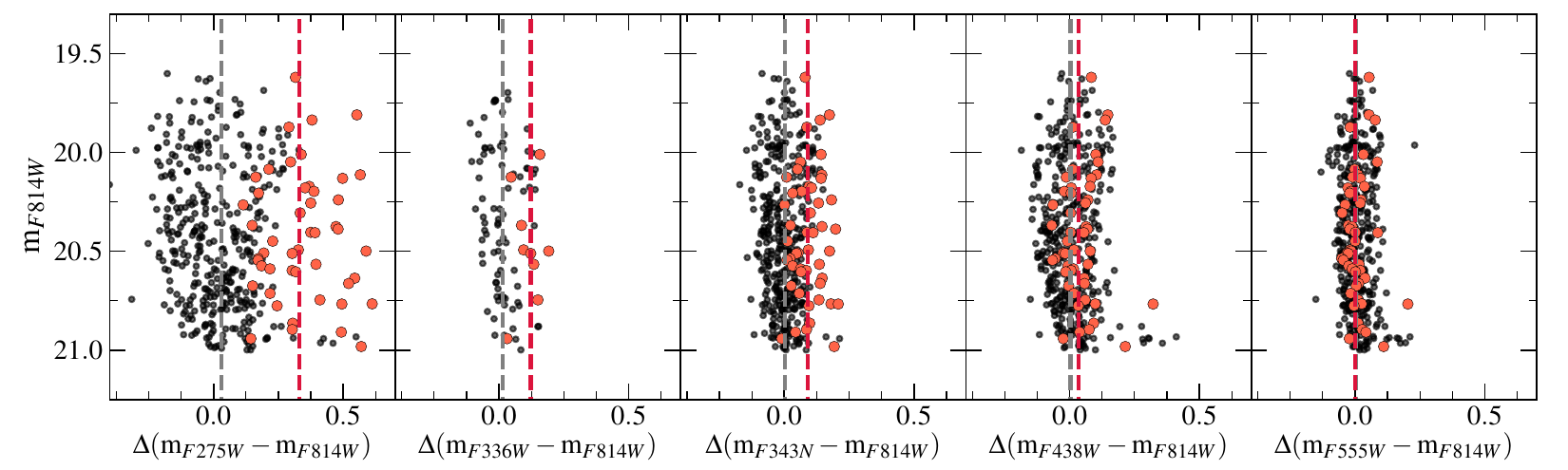}
\centering
\caption{Verticalized color distributions of the eMSTO stars of NGC 1783 in different combination of colors. The black circles represent 
eMSTO stars in the magnitude range $19.6< m_{F814W} < 21.2$, while the red circles mark the UV-dim stars selected as discussed in Section \ref{sec:uvdim}.
The gray and red vertical dashed lines mark the median values of the verticalized colors for the eMSTO and UV-dim stars, respectively.}
\label{fig:cmd_uvdim_multipanels}
\end{figure*}

\begin{figure}[h!]
\centering
\includegraphics[width=0.4\textwidth]{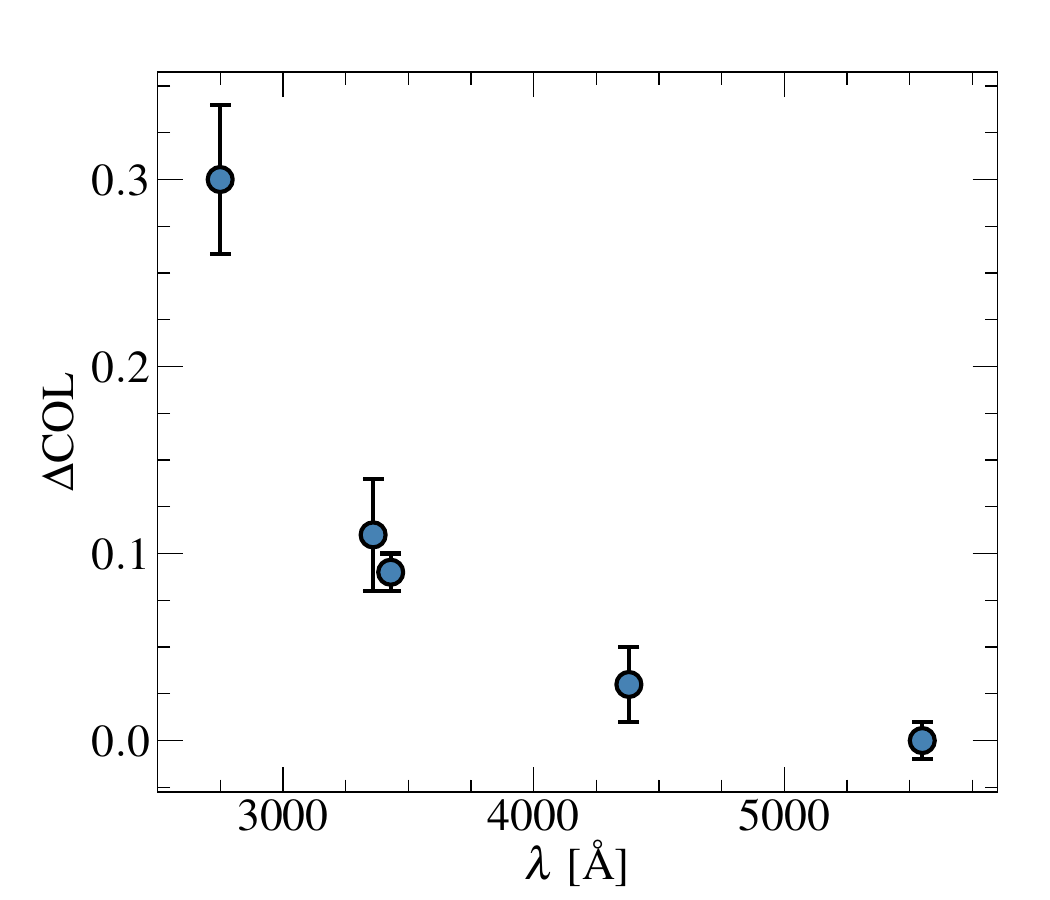}
\centering
\caption{Difference between the median of the verticalized colors of the eMSTO and UV-dim stars (see Fig. \ref{fig:cmd_uvdim_multipanels}) as a function of the wavelength. }
\label{fig:deltacol}
\end{figure}

To further test the scenario in which UV-dim stars in NGC 1783 detected in our analysis have a disk, we took advantage of the HST multi-band photometric catalog of \citet{Cadelano+22}. 
We derived the verticalized colors of the eMSTO stars following the method discussed in Section \ref{sec:uvdim} for different color combinations,
including ($m_{\rm F275W}-m_{\rm F814W}$), ($m_{\rm F336W}-m_{\rm F814W}$), ($m_{\rm F434N}-m_{\rm F814W}$), ($m_{\rm F438W}-m_{\rm F814W}$), and ($m_{\rm F555W}-m_{\rm F814W}$).
The verticalized color distributions obtained are illustrated in Fig. \ref{fig:cmd_uvdim_multipanels}, with the eMSTO and the UV-dim stars marked by black and red circles, respectively.
As expected, the UV-dim stars populate a separate region at significantly redder colors when the UV filter (F275W) is used. As the adopted filter becomes redder, the UV-dim stars progressively shift toward bluer colors, until they completely mix among the average colors of the normal eMSTO stars in the optical ($m_{\rm F814W}$, $m_{\rm F555W}-m_{\rm F814W}$) CMD.
We also computed the median values of the verticalized colors of the eMSTO and the UV-dim stars, represented in each panel by vertical, dashed gray and red lines, respectively.
It is evident that the distance between the two midlines gradually decreases until it disappears in the optical diagram.
This results is well represented in Fig. \ref{fig:deltacol}, which shows the difference between the two median values as a function of the wavelength. 
The uncertainty on the median values was estimated as the 68\% confidence interval calculated using a bootstrap algorithm with 10000 resamples.
The resulting trend suggests that UV-dim stars are likely affected 
by self-extinction, as expected in the case they host an excretion disk produced by mass loss.
Interestingly, this result is in quite good agreement with what is expected from the theoretical models presented in \citet[][see their Fig. 4, central panel, cyan curve]{d'antona+23}, which describe the presence of UV-dim stars at the level of the eMSTO in NGC 1783 by self dust absorption.\\

Finally, it is interesting to note that at the age of NGC~1783, the pulsating stars instability strip broadly intersects the location of the UV-dim stars in the ($m_{\rm F275W}-m_{\rm F438W}$ vs. $m_{\rm F814W}$) CMD, thus suggesting they might be variable stars
observed at random phase \citep[see e.g.,][]{Salinas+18}. We took advantage of the large available photometric data set to test whether UV-dim stars show any evidence of photometric variability.
To this aim we used the \texttt{variability index} \citep{steston+1996} computed by \texttt{DAOPHOT} and we performed a visual inspection of the magnitude distribution of the UV-dim stars as a function of the observation epoch for all the available filters. While admittedly the adopted dataset is not suited for a detailed variability analysis and the expected amplitudes are relatively small ($\sim 0.2$ mag) in particular at red wavelengths, we do not find any evidence of variability among the selected UV-dim stars.  

\section{Summary and discussion}
\label{sec:discussion}
 As part of a larger effort aimed at providing observational constraints on the physical mechanisms at the basis of cluster formation and early evolution (e.g., \citealt{Dalessandro+19,Dalessandro+21,Dalessandro+22,Dalessandro+24,Cadelano+22,dellacroce+23,dellacroce+24}) here we present the first extended spectroscopic analysis of the intermediate-age massive LMC cluster NGC~1783 based on deep MUSE observations providing spectra for about 2300 stars.
We derived the stellar parameters -- $T_{\mathrm{eff}}$, $\mathrm{log} \ g$, RV, and $v\mathrm{sin}i$ -- via spectral fitting following a Bayesian approach and we compared each observed spectrum with a grid of synthetic spectra by using a MCMC sampling technique. This approach offers a number of significant advantages, as it simultaneously provides reliable estimates of the best-fit values for all the parameters and the robust determinations of their associated uncertainties, taking into account the correlations among them.

In this work we focus in particular on stellar rotation. We derived $v\mathrm{sin}i$ values for about 700 stars, likely to be cluster members. In agreement with previous studies about stellar rotation in young massive clusters (e.g., \citealt{Kamann+20,kamann+23}), our analysis shows that it plays a critical role in shaping the color-magnitude distribution of stars.
In particular, we find that while sub-giant and red-giant branch stars are typically slow-rotators, the eMSTO encompasses the full range of observed rotational velocities ($0<v\sin i < 250$ km s$^{-1}$). 

Interestingly, we observe a pretty clear correlation between rotational velocities and stellar colors along the eMSTO, with rotation increasing as the color gets redder. 
However, as expected, such a correlation 
is partially blurred by the possible interplay between additional 
factors (including for example the different inclination angles and unresolved binaries) that can contribute in shaping the eMSTO morphology.

By using UV-optical HST CMDs, our analysis also confirms the presence of a sizable population ($\sim10\%$ of the eMSTOs) 
of UV-dim stars (see also \citealt{milone+23a,martocchia+23}). 
In the case of the very young (age $\sim100$ Myr) and massive cluster NGC~1850, these stars have been suggested to be likely rapidly rotating stars with decretion discs seen nearly equator-on \citep[i.e. shell stars;][]{martocchia+23}, based on direct spectroscopic observations \citep{kamann+23}. 
The idea is that UV-dim stars undergo non-negligible mass-loss through their equator triggered by their large 
rotational velocities and thus they develop an excretion disk \citep{d'antona+23}. If the disk is observed at high inclination angles with respect
to the line-of-sight then it can selectively absorb the stellar flux causing self-extinction. 
As a consequence, while shell stars are indistinguishable from eMSTOs at optical wavelengths, they 
move significantly to the red in color combinations involving UV filters.
In nice agreement with this emerging scenario, in the present analysis we find both direct and indirect evidence 
for the presence of a disk surrounding the UV-dim stars in NGC~1783. We observe stronger FeII and SiII spectral lines than those observed for standard eMSTO stars, which are typical disk-like features. We also find a monotonically decreasing distribution of the color offsets between UV-dim stars and the mean loci defined by eMSTOs, which nicely match expectations 
in the case they host a (excretion) disk.
Interestingly however, we find that UV-dim stars are typically intermediate-/low-rotators, with about $60\%$ of them having $v\mathrm{sin}i<50$ km s$^{-1}$, which is compatible with no rotation given the spectral resolution of MUSE (see Section \ref{sec:simu})
These observational evidence may suggest that UV-dim stars observed in NGC~1783 and NGC~1850 could represent distinct populations of objects at different stellar mass ranges.
Alternatively, a possible explanation involves the characterization of the magnetic field of eMSTO stars and the possibility it plays a role in shaping the observed rotation distribution.
Among O- and B-type stars, as those observed along the eMSTO of NGC~1850, only a small fraction ($\sim5\%$) is expected to host detectable magnetic fields. 
On the contrary, among A-type stars, as those populating the eMSTO of NGC~1783, up to $\sim20\%$ of stars (e.g., \citealt{abt+2000,briquet+2015}) are expected to host intermediate-strength (of up to $\sim$ 1 KG) magnetic fields  \citealt{briquet+2015}).   
Interestingly in this context, \citet{barnes+2001} suggest that (see also \citealt{matt+2004}) the development of a disk in a rotating 
magnetic star can slow-down its stellar spin (so-called disk-locking) in a relatively short timescale ranging from 5 Myr to 30 Myr, depending on the disks' rotation distribution (solid-body or differential rotation). 
This mechanism is also invoked at the level of pre-MS stars to reproduce the distribution of rotational velocities in MSTO stars in young- and intermediate-age clusters.
 However, as a cautionary note, it is worth mentioning here that the presence of magnetic fields may also inhibit the formation of circumstellar disks (e.g., \citealt{ud-Doula18})

In the framework  where differences in the magnetic field properties play a role, we can speculate that UV-dim stars in the very young cluster NGC~1850 and those in the intermediate-age cluster 
NGC~1783 may arise from the same underlying physical mechanism. In particular, in both cases these stars were originally intermediate/fast rotators that lost a fraction of their mass through their equator, mainly driven by their high spin rate. In both cases, these stars developed an excretion disk, and the fraction of them that 
are now observed equator-on are classified as UV-dim. 
In NGC~1850, where virtually all eMSTO stars lack significant magnetic fields, the presence of a disk does not substantially affect the stellar rotation velocity, resulting in UV-dim stars in this object predominantly being intermediate- to fast-rotators. Conversely, in NGC~1783, the presence of mild to relatively strong magnetic fields can decelerate the stellar spin, shifting the overall $v\mathrm{sin}i$ distribution towards intermediate/low values (see Figure \ref{fig:plots_uvdim}).

The present study tentatively suggests an evolutionary explanation for the varying properties of UV-dim stars, proposing that the fraction of rapidly-rotating UV-dim stars decreases with increasing cluster age due to a growing fraction of magnetic stars among the eMSTOs.
This idea can be directly tested through the combined analysis of the photometric and spectroscopic properties of 
UV-dim stars in clusters at different ages. In addition, high-resolution spectroscopic investigation could also provide clues 
about the existence of peculiar chemical patterns expected to be induced by the presence of magnetic fields, 
as anomalous abundances of Eu, Sr or Cr, as commonly observed in Ap-stars.

\begin{acknowledgements}
S.L. gratefully acknowledges the financial support from the project Light-on-Dark granted by MIUR through PRIN2017-2017K7REXT. 
S.M. was supported by a Gliese Fellowship at the Zentrum f\"ur Astronomie, University of Heidelberg, Germany.  
G.E. acknowledges the support provided by the Next Generation EU funds within the National Recovery and Resilience Plan (PNRR), Mission 4 - Education and Research, Component 2 - From Research to Business (M4C2), Investment Line 3.1 - Strengthening and creation of Research Infrastructures, Project IR0000034 – ``STILES - Strengthening the Italian Leadership in ELT and SKA''.
\end{acknowledgements}

%
%

\bibliographystyle{aa}
\bibliography{ngc1783}{}

\end{document}